\documentclass[10pt,journal,final]{IEEEtran}

\usepackage{graphicx}
\graphicspath{{./Figures/}}
\usepackage{subfigure} 
\usepackage{xcolor,graphicx,float} 

\usepackage{amsmath}
\usepackage{multirow}
\usepackage{amssymb}
\usepackage{algorithmic}
\usepackage{algorithm}
\usepackage{booktabs}
\usepackage{cite}
\usepackage[colorlinks,linkcolor=blue,citecolor=blue,urlcolor=black]{hyperref}
\usepackage{url}
\usepackage{titlesec}

\begin{document} 
\title{Physical Informed-Inspired Deep Reinforcement Learning Based Bi-Level Programming for Microgrid Scheduling}

\author{Yang~Li,~\IEEEmembership{Senior Member,~IEEE,} Jiankai~Gao,  Yuanzheng~Li,~\IEEEmembership{Senior Member,~IEEE}, Chen~Chen,  Sen~Li,~\IEEEmembership{Member,~IEEE},  Mohammad Shahidehpour,~\IEEEmembership{Life Fellow,~IEEE},  and Zhe Chen,~\IEEEmembership{Fellow,~IEEE}

\thanks{Yang Li is with the School of Electrical Engineering, Northeast Electric Power University, Jilin 132012, China. Yang Li is also with the Department of Civil and Environmental Engineering, The Hong Kong University of Science and Technology, Hong Kong (e-mail: liyang@neepu.edu.cn). (Corresponding author: Yuanzheng Li)}%

\thanks{Jiankai Gao is with the State Grid Datong Electric Power Supply Company, Datong 037008, China (e-mail: jiankaig@126.com).}
\thanks{Yuanzheng Li is with the School of Artifcial Intelligence and Automation,
Huazhong University of Science and Technology, Wuhan 430074, China
(e-mail: Yuanzheng\underline{~}Li@hust.edu.cn).}

\thanks{Chen Chen is with the School of Electrical Engineering, Xi’an Jiaotong University, Xi’an 710049, China (e-mail: morningchen@xjtu.edu.cn).}

\thanks{Sen Li is with the Department of Civil and Environmental Engineering,
The Hong Kong University of Science and Technology, Hong Kong
(cesli@ust.hk).}

\thanks{Mohammad Shahidehpour is with the Electrical and Computer Engineering
Department, Illinois Institute of Technology, Chicago, IL, 60616, USA. (e-mail: ms@iit.edu).}

\thanks{Zhe Chen is with the Department of Energy Technology, Aalborg University, 9220 Aalborg, Denmark (e-mail: zch@et.aau.dk).}
}

\maketitle
\markboth{}{Y.~Li \MakeLowercase{\textit{et al.}}} %

\begin{abstract}  
To coordinate the interests of operator and users in a microgrid under complex and changeable operating conditions, this paper proposes a microgrid scheduling model considering  the thermal flexibility of thermostatically controlled loads and demand response by leveraging physical informed-inspired deep reinforcement learning (DRL) based bi-level programming. To overcome the non-convex limitations of karush-kuhn-tucker (KKT)-based methods, a novel optimization solution method based on DRL theory is proposed to handle the bi-level programming through alternate iterations between levels. Specifically, by combining a DRL algorithm named asynchronous advantage actor-critic (A3C) and automated machine learning-prioritized experience replay (AutoML-PER) strategy to improve the generalization performance of A3C to address the above problems, an improved A3C algorithm, called AutoML-PER-A3C, is designed to solve the upper-level problem; while the DOCPLEX optimizer is adopted to address the lower-level problem. In this solution process, AutoML is used to automatically optimize  hyperparameters and PER improves learning efficiency and quality by extracting the most valuable samples. The test results demonstrate that the presented approach manages to reconcile the interests between multiple stakeholders in MG by fully exploiting various flexibility resources. Furthermore, in terms of economic viability and computational efficiency, the proposal vastly exceeds other advanced reinforcement learning methods.

\end{abstract}
    

\section{Introduction}  
\IEEEPARstart{W}{ith} the increasing depletion of global fossil fuels and the growing intensification of environmental pollution \cite{yu2023decarbonization}, efficient utilization of renewable energy sources has become an inevitable choice to achieve sustainable and clean energy supply  \cite{lv2019intelligent} \cite{wang2020exploring}. Because of the inherent uncertainty of renewables \cite{WANG2023120699},  renewable integration in the form of a microgrid (MG) has greatly promoted the accommodation of renewable energy resources\cite{wang2020sustainable}. Moreover, utilizing the flexibility of demand-side resources to provide ancillary services to the power system is an innovative and valuable solution that can easily be rolled out in the electricity market. Therefore, it becomes a research hotspot.

\subsection{Literature Review}  
In the reform of the electrification market, there is a competitive relationship between different stakeholders in the microgrid. Ref.\cite{HAKIMI2021117215} uses a stochastic programming algorithm for the operation planning of multiple microgrids. Ref.\cite{9696007} using model predictive control for energy management of multiple microgrids. Ref.\cite{HAN2021116830} uses a two-stage stochastic strategy for energy trading in multiple microgrids. However, the aforementioned studies mainly coordinate energy supply entities and ignore the interaction of interests between the demand side and energy manager. Moreover, the statistical results show that thermostatically controlled loads (TCLs) consume about 20$\%$ of the electricity in the total loads in the United States such that TCLs have great potential in improving the operational flexibility of microgrids (MGs). How to leverage the flexibility of thermostatically controlled loads to participate in demand response (DR) to coordinate multi-stakeholders has been  receiving increasing attention.

Usually, such problems are modeled as a bi-level programming based on the following reasons: there is a hierarchical relationship between the decisions of two stakeholders, and a decision of each party affects that of the other party. Regarding the solution of bi-level optimization problems, it is usually divided into model-based and data-driven methods. There has been various research on model-based approaches here.  Ref. \cite{LI2021113996} uses karush-kuhn-tucker (KKT) conditions to convert the bi-level issue into a mathematical programming with equilibrium constraints (MPEC) for the solution. Ref. \cite{li2021coordinating} proposes a dynamic pricing mechanism to coordinate the benefits of community integrated energy system (CIES) and electric vehicle charging station (EVCS). Ref. \cite{peng2021hybrid} uses a distributed algorithm to find the optimal solution of the model. Ref. \cite{liu2019heat} adopts differential evolution combined with CPLEX to solve a bi-level model of combined heat-and-power (CHP) unit owners and industrial users. Unfortunately, since the problem of coordinating the interests of different stakeholders in a MG usually involves unknown or uncertain information, the above model-based methods suffer from the following challenges: (1) the performance of the model-based method depends on the accuracy of the model used and its parameters; (2) in traditional model-based methods, the introduction of multi-stakeholder and thermostatically controlled loads can make it more difficult to deal with uncertainty; (3) the above-mentioned iterative algorithm is difficult to guarantee convergence, which requires certain simplifications and assumptions. And the time-consuming and resource-intensive iteration of the algorithm is unrealistic; (4) when treating with a MG scheduling model with nonlinearity, the equivalent KKT condition cannot address non-convex problems. Therefore, how to develop an effective method to address these challenges is a key issue for MG optimal scheduling.

To resolve the aforementioned difficulties, we leverage a deep reinforcement learning (DRL) algorithm in a data-driven strategy that balances the interests of numerous stakeholders. In this case, a deep neural network may perform effective model regression without depending on exact formulas of mathematics by  autonomously extracting characteristics from the input. In contrast to conventional optimization techniques, reinforcement learning (RL) brings up a new approach to address non-convex problems since it is ideal for instantaneous decision-making within complicated and variable operating circumstances. At present, there are literatures that apply RL to solve the optimal scheduling of MG. Ref. \cite{9451164} proposes a DRL-based privacy-preserving load control scheme to reduce the operating cost of MG. In Ref. \cite{9244070}, a multi-agent safe RL method is proposed for power/energy management systems of multiple MGs. Besides, Ref. \cite{8331897} utilizes RL for distributed energy management. Ref. \cite{9509287} develops an online operation strategy for MG based on RL. Ref. \cite{8839066} employs RL for energy trading in MG.  However, the above studies only focus on the energy management optimization of microgrid with a single stakeholder.

To characterize the interaction of different entities in the energy trading process, some studies have used RL to provide managers with decision-making through iteratively interacting with the environment to learn. Ref. \cite{8789677} presents a bi-level RL framework based on price. Ref. \cite{8769895} proposes a RL for energy management of multiple MGs to reduce the peak-to-average ratio on the demand side. Ref. \cite{ZHU2021117107} proposes a multi-agent reinforcement learning (MARL) method to enable virtual MGs to dynamically update their bidding strategies according to previous market settlements. As far as the above research is concerned, while RL is used to optimize multi-stakeholder entities, it does not fully exploit the high generalization ability and decision-making efficiency of DRL. In this regard, an asynchronous advantage actor-critic (A3C) algorithm combining value and policy iteration is successfully applied to power systems \cite{9099902}. To coordinate the interests of different entities in MGs, we propose a method that combines automated machine learning (AutoML) and prioritized experience replay (PER) with A3C to further improve the generalization performance of DRL and address the shortcomings of existing A3C.  Furthermore, the proposed method has the following significant advantages over other RL: the advantage function to judge the quality of the action; the asynchronous training framework that can significantly improve learning speed; AutoML is used to automatically optimize hyperparameters, and PER improves learning efficiency and quality by extracting the most valuable samples.

\begin{table}
  \centering
  \caption{Comparison of the proposed method with recent related works in MG scheduling }\label{tab1}
    \setlength{\tabcolsep}{0.1mm}{
    \begin{tabular}{c|c|c|c|c|c|c|c}
    \hline
    \multirow{3}{*}{Ref} &  \multicolumn{2}{c|}{\multirow{2}{*}{Stakeholders}} & \multirow{3}{*}{\begin{tabular}[c]{@{}c@{}} Inter-level \\ interaction  \end{tabular}}   & \multicolumn{2}{c|}{\multirow{2}{*}{Optimization of each level}}   & \multirow{3}{*}{DR} & \multirow{3}{*}{TCL} \\ & \multicolumn{2}{c|}{}                   &                    & \multicolumn{2}{c|}{}                   &                    &\\
\cline{2-3}\cline{5-6}          & upper level & lower level &       & upper level & lower level &       &  \\   \hline
  
    \cite{LI2021113996}    & {\begin{tabular}[c]{@{}c@{}}integrated \\energy \\operator\end{tabular}} & users & {\begin{tabular}[c]{@{}c@{}} $\Xi $ \end{tabular}} & CPLEX & $-$    & $\surd$     & × \\ \hline
    \cite{li2021coordinating}    & CIES  & EVCS  &  {\begin{tabular}[c]{@{}c@{}}$\Lambda$\end{tabular}}  & CPLEX & CPLEX & $\surd$     & × \\ \hline
    \cite{peng2021hybrid}    & {\begin{tabular}[c]{@{}c@{}}integrated \\energy \\operator\end{tabular}} & prosumers & {\begin{tabular}[c]{@{}c@{}} $\Lambda$ \end{tabular}} & {\begin{tabular}[c]{@{}c@{}} distributed \\algorithm\end{tabular}} & {\begin{tabular}[c]{@{}c@{}} distributed \\algorithm\end{tabular}}  & $\surd$     & × \\ \hline
    \cite{liu2019heat}    & {\begin{tabular}[c]{@{}c@{}} CHP unit \\ owners\end{tabular}}  & {\begin{tabular}[c]{@{}c@{}} industrial\\users\end{tabular}} & {\begin{tabular}[c]{@{}c@{}} $\Lambda$ \end{tabular}} & {\begin{tabular}[c]{@{}c@{}} differential\\evolution\end{tabular}} & CPLEX & $\surd$     & × \\  \hline
    \cite{8789677} & {\begin{tabular}[c]{@{}c@{}} cooperative \\ agent \end{tabular}}  & {\begin{tabular}[c]{@{}c@{}} MG control \\center\end{tabular}} & {\begin{tabular}[c]{@{}c@{}}$\Lambda$ \end{tabular}} & {\begin{tabular}[c]{@{}c@{}} RL\end{tabular}} & RL & ×     & × \\  \hline 
     
    \cite{8769895} & {\begin{tabular}[c]{@{}c@{}} each MG \\ operator \end{tabular}}  & {\begin{tabular}[c]{@{}c@{}} distribution \\system operator\end{tabular}} & {\begin{tabular}[c]{@{}c@{}}$\Lambda$ \end{tabular}} & {\begin{tabular}[c]{@{}c@{}} GAMS\\/CPLEX\end{tabular}} & RL & $\surd$     & × \\  \hline
    \cite{ZHU2021117107} & {\begin{tabular}[c]{@{}c@{}} distribution \\ system operator \end{tabular}}  & {\begin{tabular}[c]{@{}c@{}}virtual \\MGs\end{tabular}} & {\begin{tabular}[c]{@{}c@{}}$\Lambda$ \end{tabular}} & {\begin{tabular}[c]{@{}c@{}} MARL\end{tabular}} & MARL & ×     & × \\  \hline
    \textbf{{\begin{tabular}[c]{@{}c@{}} This\\paper\end{tabular}}} & \textbf{ {\begin{tabular}[c]{@{}c@{}} MG\\operator\end{tabular}}} & \textbf{users} & \textbf{$\Upsilon$} & \textbf{{\begin{tabular}[c]{@{}c@{}} improved \\ A3C \end{tabular}}} & \textbf{docplex} & \textbf{$\surd$}     & \textbf{$\surd$} \\  \hline
    \end{tabular}%
  }
\end{table}%

For summarizing the uniqueness of the proposed approach, the comparison between the proposal and the recent related works in the field of MG scheduling is shown in Table \ref{tab1}. Regarding inter-level interactions, symbol $\Xi $ refers to use the KKT to convert bi-level programming into a single-level MPEC; while symbols $\Lambda$ and $\Upsilon$ are respectively traditional- and our DRL-based alternate iterations between two levels, where the decision of the upper level is passed to the lower level in the same way. 
In terms of bi-level iterative methods, the differences between our approach and $\Lambda$ are as follows: $\Lambda$ indicates that the lower-level strategy is passed to the upper-level objective function as a decision variable to be optimized, 
while $\Upsilon$ denotes that the lower-level strategy is passed to the upper-level environment as a state, and then guides the update of reward. The reason we take this approach is that DRL is better suited to solve the unknown implicit black box problems, which pass low-level decisions as state to the upper environment and then utilize the high generalizability of DRL to formulate more long-term benefits for  operator and users. However, to the authors' best knowledge, so far there is little study that utilizes the above-mentioned deep reinforcement learning theory-based bi-level optimization to coordinate the interests of multi-stakeholders in MG scheduling.

\subsection{Contribution of This Paper}
The main contributions of this paper can be highlighted as: 
\begin{enumerate}

    \item In order to solve the problem of coordination of interests of different stakeholders, a bi-level scheduling model of microgrid operator and users is constructed by using the flexibility of thermostatically controlled loads and demand response.

    \item To overcome the non-convex limitations of KKT-based methods and the shortcomings of traditional bi-level iterative methods those are unable to address complex and changeable scheduling problems, we treat the lower-level problem as a black box and use its decision variables as the upper-level state to deal with bi-level programming, which is a new solution methodology.

    \item By combining A3C and AutoML-PER strategy, a DRL algorithm, called AutoML-PER-A3C, is designed to improve the generalization performance of A3C, so that the agent can learn the optimal scheduling policy for coordinating the interests of multiple entities from a long-term perspective by feeding back the state from the lower level to the upper level. Moreover, the rationality of the agent's policies will be better explained by the physical informed-inspired reward.

\end{enumerate}

\section{Microgrid Physical Model}\label{The model}
\begin{figure}[t]
    \centering
    \includegraphics[width=2.9in,height=2.0in]{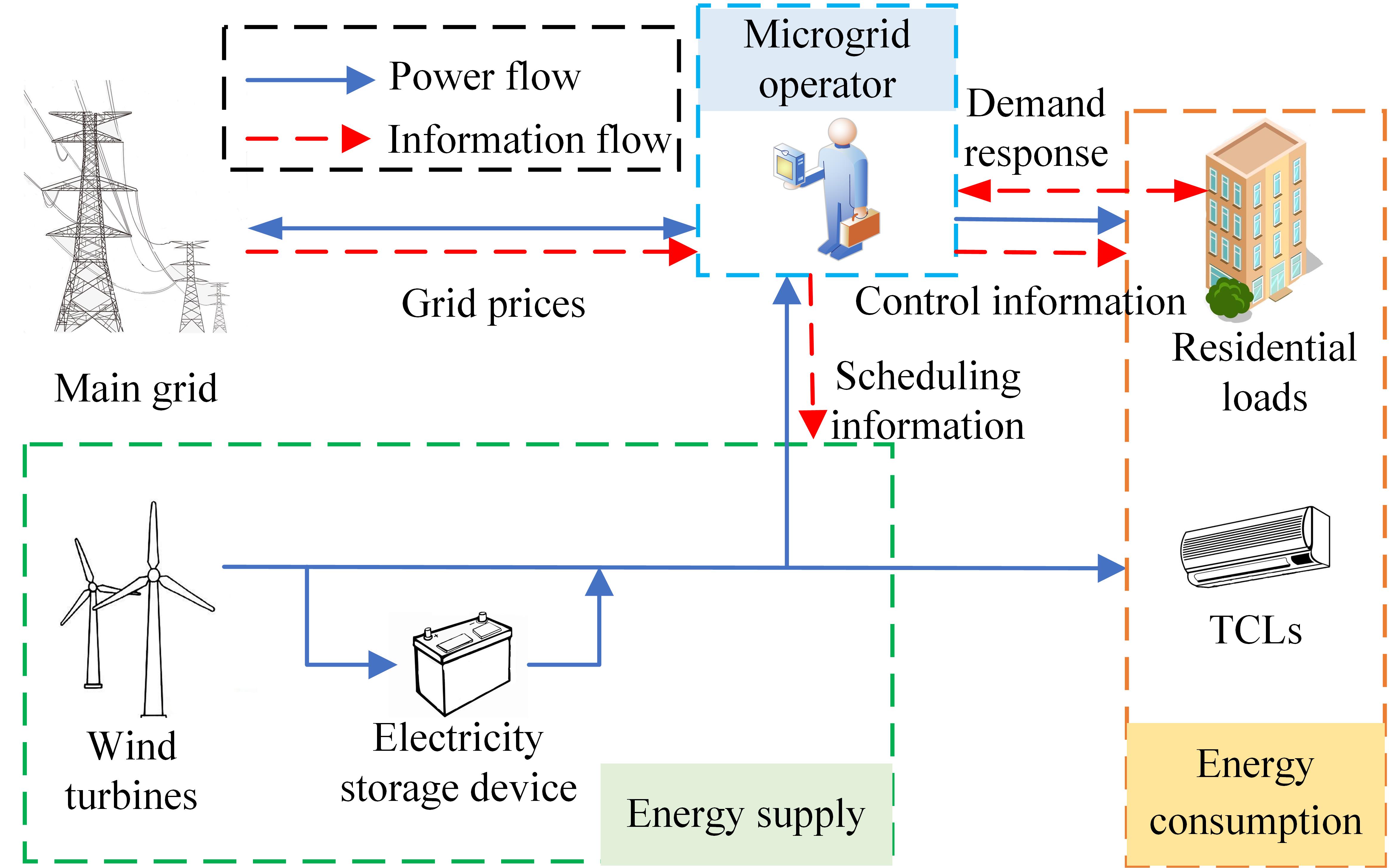}
    \caption{Schematic diagram of the studied microgrid.}
    \label{microgrid}
\end{figure}

To clearly show its physical model, Fig. \ref{microgrid} illustrates a schematic diagram of a MG, which mainly consists of wind turbines (WT) units, electricity storage device (ESD), TCLs and residential price response loads.

\subsection{Distributed Generation}
The MG being studied has a WT distributed production. Relying on the concepts of data-driven scheduling, this study effectively uses real wind generation data for the following study, instead than estimating WT power generates in the manner of an explicit statement.

\subsection{Load Model}
\subsubsection{Thermostatically Controlled Load}
TCLs generally participate in load regulation in the form of an aggregated cluster. Note that the electricity cost of TCLs is charged at the price of wind power (rather than users' real-time electricity price) to compensate for the decreased user comfort levels. At each time step $i$, the switching action of the $n$-th TCL  is a binary variable:
{\setlength\abovedisplayskip{2pt}
\setlength\belowdisplayskip{2pt}
\begin{equation}\label{TCL1}
{s_{i,n}} \in \left\{ {0,1} \right\}
\end{equation} } 

Each TCL is equipped with a backup controller, which acts as a filter to control the switch action ${s_{i,n}}$. The actions of the backup controller depend on the switch action and  indoor air temperature. When the TCLs users' comfort is decreased, the backup controller will activate the TCLs, thereby ensuring the comfortable  indoor temperature. Therefore, the actual physical actions of the backup controller \cite{7180409} are described as follows: 
{\setlength\abovedisplayskip{2pt}
\setlength\belowdisplayskip{2pt}
\begin{equation}\label{TCL3}
B({T_n}(t),{s_{i,n}}) = \left\{ \begin{array}{l}
1{\rm{ ,  \quad    if \quad }}{T_n}(t) < {T_{LB,n}}(t)\\
{s_{i,n}}{\rm{, if \quad }}{T_{LB,n}}(t) \le {T_n}(t) \le {T_{UB,n}}(t)\\
0{\rm{  ,  \quad  if \quad }}{T_n}(t) > {T_{UB,n}}(t)
\end{array} \right.{\rm{  }}\forall t
\end{equation} }where ${T_{UB,n}}(t)$ and ${T_{LB,n}}(t)$ respectively denote the upper and lower bounds of the comfort range set by the TCL users,  ${T_n}(t)$ is the indoor air temperature, and $B({T_n}(t),{s_{i,n}})$ is a binary variable that reflects the start and stop actions of the backup controller. In this paper, the second-order equivalent thermal parameters (ETP) model \cite{7303968} is used to depict the dynamic change process of indoor temperature and TCLs actions. The details are as follows:

{\setlength\abovedisplayskip{2pt}
\setlength\belowdisplayskip{2pt}
\begin{footnotesize}
\begin{equation}
\begin{aligned}\label{TCL4}   
{\dot T_n}\left( t \right) &= \frac{1}{{{C_{air,n}}}}[{T_{out}}(t) - {T_n}(t)] + \frac{1}{{{C_{m,n}}}}[{T_{m,n}}(t) - {T_n}(t)]\\
&+ {P_{TCL,{\rm{n}}}}B({T_n}(t),{s_{i,n}}) + {Q_{air}}\quad \forall t
\end{aligned}
\end{equation}
\end{footnotesize}}{\setlength\abovedisplayskip{2pt}
\setlength\belowdisplayskip{2pt}
\begin{equation}\label{TSL5}
   {\dot T_{m,n}}\left( t \right) = \frac{1}{{{C_m}}}[{T_n}(t) - {T_{m,n}}(t)]\quad \forall t
\end{equation} }where ${T_{m,n}}(t)$ and ${T_{out}}(t)$ represent the building $m$ temperature and outdoor ambient temperature, respectively; ${P_{TCL,{\rm{n}}}}$ is the total power of all TCLs;  ${C_{air,n}}$ and ${C_{m,n}}$ are the equivalent heat capacity of the air and the building mass temperature; and ${Q_{air}}$ is the heat flow of the air. Moreover, the indoor air temperature within the users' comfort range is represented by the state of charge $SO{C_{TCL,n}}$ , which is 
{\setlength\abovedisplayskip{2pt}
\setlength\belowdisplayskip{2pt}
\begin{equation}\label{TCL6}
SO{C_{TCL,n}} = \frac{{{T_n}(t) - {T_{LB,n}}(t)}}{{{T_{UB,n}}(t) - {T_{LB,n}}(t)}} \quad \forall t
\end{equation} }

When the indoor temperature falls into the comfortable range, the MG operator transmits an energy distribution level control signal $\iota $ corresponding to the TCL action to a TCL aggregator. And the aggregator determines the switching action based on the priority of all TCL states of charge and the level of energy distribution $\iota $. The lower the $SO{C_{TCL,n}}$ value, the TCL will act first.
\subsubsection{Non-TCLs residential loads }
The residential load \cite{nakabi2021deep} is divided into fixed load and flexible load, where the latter refers to time-shiftable load. The actual residential load in period $t$ is as follows:
{\setlength\abovedisplayskip{2pt}
\setlength\belowdisplayskip{2pt}
\begin{equation}\label{RL1}
{P_{load,t}} = {P_{basic,t}} - {P_{TSL,t}} + {P_{PBL,t}}\quad \forall t
\end{equation} }{\setlength\abovedisplayskip{2pt}
\setlength\belowdisplayskip{2pt}
\begin{equation}\label{RL2}
{P_{TSL,t}} = {P_{basic,t}}{\sigma _t}{\varsigma _t}\quad \forall t
\end{equation} }{\setlength\abovedisplayskip{2pt}
\setlength\belowdisplayskip{2pt}
\begin{equation}\label{RL4}
{P_{PBL,t}} = \sum\limits_i^{t - 1} {{\omega _{i,t}}{P_{TSL,t}}}\quad \forall t
\end{equation} }where ${P_{load,t}}$ denotes the actual load value in period $t$; ${P_{basic,t}}$ is the fixed  load value; ${P_{TSL,t}}$ represents the load transferred ; ${P_{PBL,t}}$ is the transferred load paid back;  ${\sigma _t}$ is the sensitivity factor, which describes the percentage of load change as price fluctuates;  ${\varsigma _t}$ is the price level, which determines the residential load actions;  ${\omega _{i,t}} \in \left\{ {0,1} \right\}$ 
denotes the executive decision of the load transferred from  time step $i$ in period $t$, which is determined by the probability $P$ that the load transfer will be reimbursed. 
{\setlength\abovedisplayskip{2pt}
\setlength\belowdisplayskip{2pt}
\begin{equation}\label{RL5}
P = clip\left( {\left. {\frac{{ - {\varsigma _t}(sign({P_{TSL,t}}))}}{2} + \frac{{t - i}}{\eta },0,1} \right)} \right. \quad \forall t
\end{equation} }{\setlength\abovedisplayskip{2pt}
\setlength\belowdisplayskip{2pt}
\begin{equation}\label{RL6} 
clip(\emph{z},0,1) = \left\{ \begin{array}{l}
0, \quad{\rm{   if\quad \emph{z} < 0}}\\
{\rm{\emph{z}, \quad if \quad 0}} \le {\rm{\emph{z}}} \le {\rm{1}}\\
{\rm{1, \quad if\quad \emph{z} > 1}}
\end{array} \right.
\end{equation} }where $clip()$ is used to limit the load transfer probability;  $\eta $ is the patience value, which describes the number of hours when the transferred load is repaid. ${\omega _{i,t}} = 1$ indicates that the load is transferred; otherwise,  ${\omega _{i,t}} = 0$ means that the transferred load will not be repaid. Note that when the time elapsed from time step $i$ is closer to the maximum patience value, 
loads with a higher probability will be transferred. 

Moreover, see Appendix A for the ESD model.

\section{Formulation of Microgrid Scheduling}\label{Problem formulation}
This section first introduces a dynamic pricing mechanism in the upper level, and then formulates the problems of the upper- and lower- levels in detail. Note that the upper-level operator releases electricity price $\lambda _t$ to lower-level users, and users adjust energy consumption strategies $P_{TCL,n}$ based on the electricity price.  
\subsection{Dynamic Pricing Mechanism}
In this study, the price is set by the operator based on the market price, and takes into account price fluctuations determined by the user behavior habits. According to the energy consumption strategy, the price set by the operator fluctuates around the market price ${\lambda _{market}}$, the difference between the daily average price and the market price ${{\lambda }_{average}}-{{\lambda }_{market}}$ does not exceed ${\rho}$ of the market price. To account for the impact of future earnings, we set the price level ${\varsigma _t}$ based on learned strategies in Section \ref{Model Solving}. The specific pricing is as follows.
{\setlength\abovedisplayskip{2pt}
\setlength\belowdisplayskip{2pt}
\begin{equation}\label{DPM1}
{\lambda _t} = {\lambda _{market}} + {\varsigma _t}\kappa \quad \forall t
\end{equation} }{\setlength\abovedisplayskip{2pt}
\setlength\belowdisplayskip{2pt}
\begin{equation}\label{DPM2}
\frac{{{\lambda _{average}} - {\lambda _{market}}}}{{{\lambda _{market}}}} < {\rho}
\end{equation} }where constant $\kappa=1.5$  represents  the change in the price, the threshold $\rho=2.9${\%}  stands for the upper limit of the ratio of the difference between the daily average price and the market price to the market price, ${\lambda _t}$ is the electricity price released by the operator in period $t$.

\subsection{Design of Bi-Level Model of Microgrid}
In the bi-level transaction process, the operator releases electricity prices to lower-level users, and users adjust energy consumption strategies based on the electricity price. Furthermore, the decision variable at the upper level is the price strategy ${\lambda _t}$ and the lower level is the energy consumption strategy ${P_{TCL,n}}$.  

\subsection{The Upper-Level Model}
\subsubsection{ Objective Function}
The upper level aims to maximize the net profits of MG operator. The MG income comes from electricity sale to the main grid, residential loads and TCLs; while the operating cost includes the cost of buying electricity from the grid, the compensation cost of shifting loads, and the power transmission cost when the MG trades electricity with the grid. Considering the compensation cost of the transferred load, the objective function ${F_1}$ of the upper level is
{\setlength\abovedisplayskip{2pt}
\setlength\belowdisplayskip{2pt}
\begin{equation}\label{OF1}
\max {F_1} = {f_{income}} - {f_{cost}}
\end{equation} }{\setlength\abovedisplayskip{2pt}
\setlength\belowdisplayskip{2pt}
\begin{footnotesize}
\begin{equation}
\begin{aligned}\label{OF2}
{f_{income}} &= {\lambda _t}\sum\limits_l^{loads} {{P_{load,t}} + {\lambda _{gen}}\sum\limits_n^{TCLs} {{P_{TCL,n}}B({T_n}(t),{s_{i,n}})} }\\
&+ {\lambda _{sell,t}}{P_{sell,t}}
\end{aligned}  
\end{equation} 
\end{footnotesize}}{\setlength\abovedisplayskip{2pt}
\setlength\belowdisplayskip{2pt}
\begin{footnotesize}
\begin{equation}
\begin{aligned}\label{OF3}
{f_{cost}} &= \zeta \sum\limits_k^{TSL} {\max ({P_{TSL,t}},0) + } {P_{buy,t}}({\lambda _{buy,t}}\\ 
&+ {\mu _{import}}) + {\mu _{export}}{P_{sell,t}}
\end{aligned}
\end{equation} 
\end{footnotesize}}where ${f_{income}}$ and ${f_{cost}}$ are the total income of the operator and the MG operating cost;  ${\lambda _{gen}}$ denotes the cost price of power generation of the WT;  ${\lambda _{buy,t}}$ and ${\lambda _{sell,t}}$  represent the prices of electricity purchased and sold when transacting with the main grid; ${P_{buy,t}}$ and ${P_{sell,t}}$ are the electricity purchased and sold when transacting with the main grid;  ${\mu _{import}}$ and ${\mu _{export}}$ are the import and export transmission costs for purchasing from and selling to the main grid; $\zeta$ is the unit compensation coefficient of the transferred load; $l$, $n$ and $k$ represent the $l$-th residential load, the $n$-th TCL and the $k$-th time-shiftable load.        % 

\subsubsection{Constraints }
\paragraph{Price constraint}
To ensure the revenue of the operator, the electricity prices issued by the operator are set by
{\setlength\abovedisplayskip{2pt}
\setlength\belowdisplayskip{2pt}
\begin{equation}\label{jiageyueshu}
{\lambda _{\min }} \le {\lambda _t} \le {\lambda _{\max }}{\rm{  }}\quad \forall t
\end{equation} }where ${\lambda _{\min }}$ and ${\lambda _{\max }}$ represent the minimum and maximum electricity prices, and ${\lambda _t}$ is the price issued in period $t$.

\paragraph{Network constraints}
In this study, without considering the network losses, the sum of the power output of the WT, the power output of the ESD and the electrical power supplied by the grid to the MG system is equal to the sum of the load power, the power delivered to the grid and the electrical power consumed by the ESD. And to reasonably regulate the power of nodes in the MG, the power balance constraint is expressed as
{\setlength\abovedisplayskip{2pt}
\setlength\belowdisplayskip{2pt}
\begin{footnotesize}
\begin{equation}
\begin{aligned}\label{Network1}
{P_{WT,t}} + k_1^t{P_{buy,t}} &+ k_3^t{P_{dc,t}} - \sum\limits_l^{loads} {{P_{load,t}} - k_2^t{P_{sell,t}}} \\& -\sum\limits_n^{TCLs} {{P_{TCL,n}}B\left( {{T_n}\left( t \right),{s_{i,n}}} \right)} - k_4^t{P_{ch,t}} = 0  \quad\forall t 
\end{aligned}
\end{equation}
\end{footnotesize}}{\setlength\abovedisplayskip{2pt}
\setlength\belowdisplayskip{2pt}
\begin{equation}\label{Network2}
\left\{ \begin{array}{l}
k_1^t,k_2^t \in \left\{ {0,1} \right\}\\
k_1^t + k_2^t \le 1
\end{array} \right.{\rm{  }}\forall t
\end{equation} }{\setlength\abovedisplayskip{2pt}
\setlength\belowdisplayskip{2pt}
\begin{equation}\label{Network3}
\left\{ \begin{array}{l}
k_3^t,k_4^t \in \left\{ {0,1} \right\}\\
k_3^t + k_4^t \le 1
\end{array} \right.{\rm{  }}\forall t
\end{equation} }where $k_1^t$, $k_2^t$ as well as $k_3^t$, $k_4^t$ respectively represent the binary decision variables for buying and selling electricity as well as charging and discharging powers in period $t$, and ${P_{WT,t}}$ denotes the power of the WT.

To prevent line power from exceeding the rated range, the capacity limits are as follows
{\setlength\abovedisplayskip{2pt}
\setlength\belowdisplayskip{2pt}
\begin{equation}\label{Network4}
{P_{ab,\min }} \le {P_{ab,t}} \le {P_{ab,\max }}
\end{equation} }where ${P_{ab,t}}$ is the apparent power flow between nodes $a$ and $b$ at time $t$, ${P_{ab,\min }}$ and ${P_{ab,\max }}$ are the minimum and maximum apparent power flow between nodes $a$ and $b$.

In addition, see Appendix B for the ESD constraints.

\subsection{The Lower-level Model}
\subsubsection{ Objective Function}
The lower level aims to seek the minimization of the electricity cost ${F_2}$ of the residential loads and TCLs, which represents the interest of the users and is formulated by
{\setlength\abovedisplayskip{2pt}
\setlength\belowdisplayskip{2pt}
\begin{equation}\label{xiacengmubiao}
\min {F_2} = {\lambda _t}\sum\limits_l^{loads} {{P_{load,t}} + {\lambda _{gen}}\sum\limits_n^{TCLs} {{P_{TCL,n}}B({T_n}(t),{s_{i,n}})} } 
\end{equation} }

\subsubsection{Constraints}
\paragraph{Power balance constraint} % 
To reasonably adjust the energy utilization of the ESD and main grid, the power balance constraint can be expressed as
\setlength{\abovedisplayskip}{2pt}
\setlength{\belowdisplayskip}{2pt}
{\footnotesize
\begin{equation}
\begin{aligned}\label{llyueshu1} 
P_{WT,t} &+ k_1^t P_{buy,t} + k_3^t P_{dc,t} = \sum_{l}^{loads} (P_{load,t} + k_2^t P_{sell,t}) \\
&+ \sum_{n}^{TCLs} (P_{TCL,n} B(T_n(t), s_{i,n})) + k_4^t P_{ch,t} \quad \forall t
\end{aligned}
\end{equation}
}{\setlength\abovedisplayskip{2pt}
\setlength\belowdisplayskip{2pt}
\begin{equation}\label{llyueshu2}
\left\{ \begin{array}{l}
k_1^t,k_2^t \in \left\{ {0,1} \right\}\\
k_1^t + k_2^t \le 1
\end{array} \right.{\rm{  }}\forall t
\end{equation} }{\setlength\abovedisplayskip{2pt}
\setlength\belowdisplayskip{2pt}
\begin{equation}\label{llyueshubuchong}
\left\{ \begin{array}{l}
k_3^t,k_4^t \in \left\{ {0,1} \right\}\\
k_3^t + k_4^t \le 1
\end{array} \right.{\rm{  }}\forall t
\end{equation} }%

In addition, to capture the non-convex operation characteristics of the discrete level of electricity purchase and sale when trading with the grid, we use a binary decision variable constraint (\ref{llyueshu2}) to ensure that buying and selling electricity will not happen at the same time. Unfortunately, the KKT method is not directly capable of addressing non-convex optimization and the convex transformation of large-scale non-convex problems is complicated. More importantly, this study focuses on microgrid scheduling under uncertain operating conditions, while using the KKT method requires the certain information. For these reasons, we propose a new DRL method to solve the bi-level optimization problem in a recursive manner. The specific solution method will be introduced in detail in Section \ref{Model Solving}.

\paragraph{Power constraints of TCLs}
According to the second-order ETP model of TCLs, the TCL power  directly affects the change of indoor temperature. Specifically, we set the following TCL power constraints according to the power change of TCL:
{\setlength\abovedisplayskip{2pt}
\setlength\belowdisplayskip{2pt}
\begin{equation}\label{llyueshu7}
{P_{TCL,\min }} \le {P_{TCL,n}} \le {P_{TCL,\max }}{\rm{  }}\quad \forall t
\end{equation} }where ${P_{TCL,\min }}$ and ${P_{TCL,\max }}$ are the minimum and maximum values of the power of TCLs.

\paragraph{Constraints on transaction electricity with the grid}
To prevent excessive purchase and sale of electricity, the transaction electricity that the operator buys electricity from or sells electricity to the main grid in period $t$  is set by
{\setlength\abovedisplayskip{2pt}
\setlength\belowdisplayskip{2pt}
\begin{equation}\label{llyueshu8}
{\rm{0 \le }}{{\rm{P}}_{buy,t}} \le {P_{\max }}\quad \forall t
\end{equation} }{\setlength\abovedisplayskip{2pt}
\setlength\belowdisplayskip{2pt}
\begin{equation}\label{llyueshu9}  
{\rm{0 \le }}{{\rm{P}}_{sell,t}} \le {P_{WT,t}}\quad \forall t
\end{equation} }where ${P_{\max }}$ is the maximum value of ${\rm{P}}_{buy,t}$.

Furthermore, see Appendix B for the ESD constraints.

\section{Model Solving}\label{Model Solving}
In this section, the proposed AutoML-PER-A3C and DOCPLEX approaches for solving the built upper- and lower- models are described in detail. 

\subsection{Automated Machine Learning}
Normally, selecting the right hyperparameters and designing a neural network is an exercise in trial and error in conventional machine learning methods. However, the problem is that these chores are often menial and tiresome.
To this end, we utilize a complex control framework to run the machine learning model, so it can learn the suitable parameters and settings without human involvement\cite{hutter2019automated}, \cite{awad2015machine}, \cite{he2021automl}.

\begin{figure}[t]
    \centering
    \includegraphics[width=3.1in,height=1.5in]{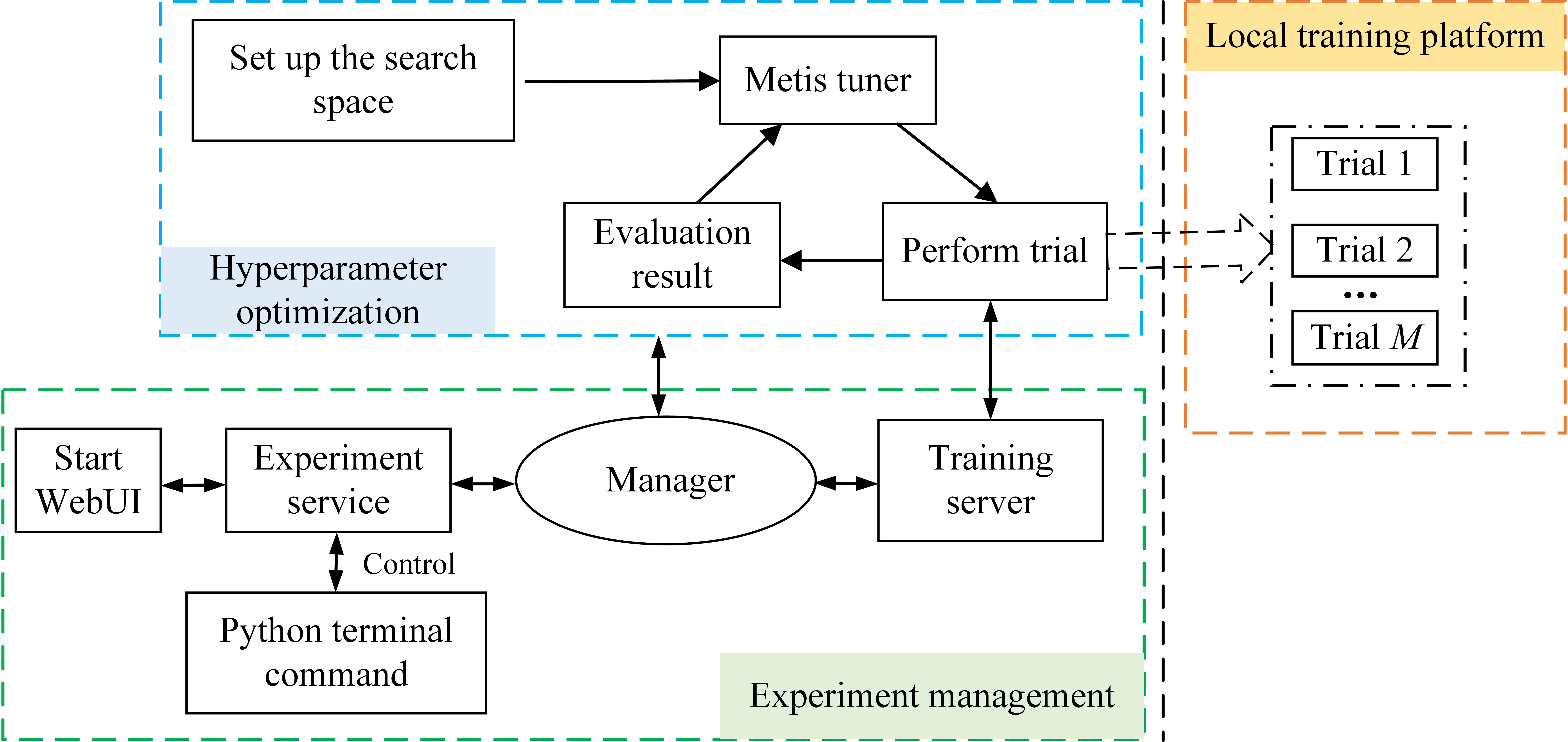}
    \caption{DRL hyperparameter optimization based on the AutoML.}
    \label{nni}
\end{figure}

It is well-known that optimizing the DRL algorithm's hyperparameters is a difficult undertaking. In this work, optimal hyperparameter combinations for DRL are determined using the widely employed AutoML technology, the architecture of which is shown in Fig. \ref{nni}. In order to tune A3C's hyperparameters, this study employs the Metis Tuner \cite{li2018metis} method.  Besides, AutoML calculates the ideal hyperparameters for the A3C by using the Metis to forecast the next trial rather than random guesswork.

\subsection{AutoML-PER-A3C Methodology}
The markov decision process (MDP) is a useful framework for describing DRL problems in general. In detail, MDP is composed of five key elements $\left\{ {s,a,\rho ,r,\gamma }\right\}$, where $s$ is the state,  $a$ represents the action,  $\rho $ denotes the state transition probability matrix,  $r$ is the reward from states $s{}_{t - 1}$ to ${s_t}$ , and $\gamma$ is the discount factor.

\begin{figure}[t]
    \centering
    \includegraphics[width=3.5in,height=3.8in]{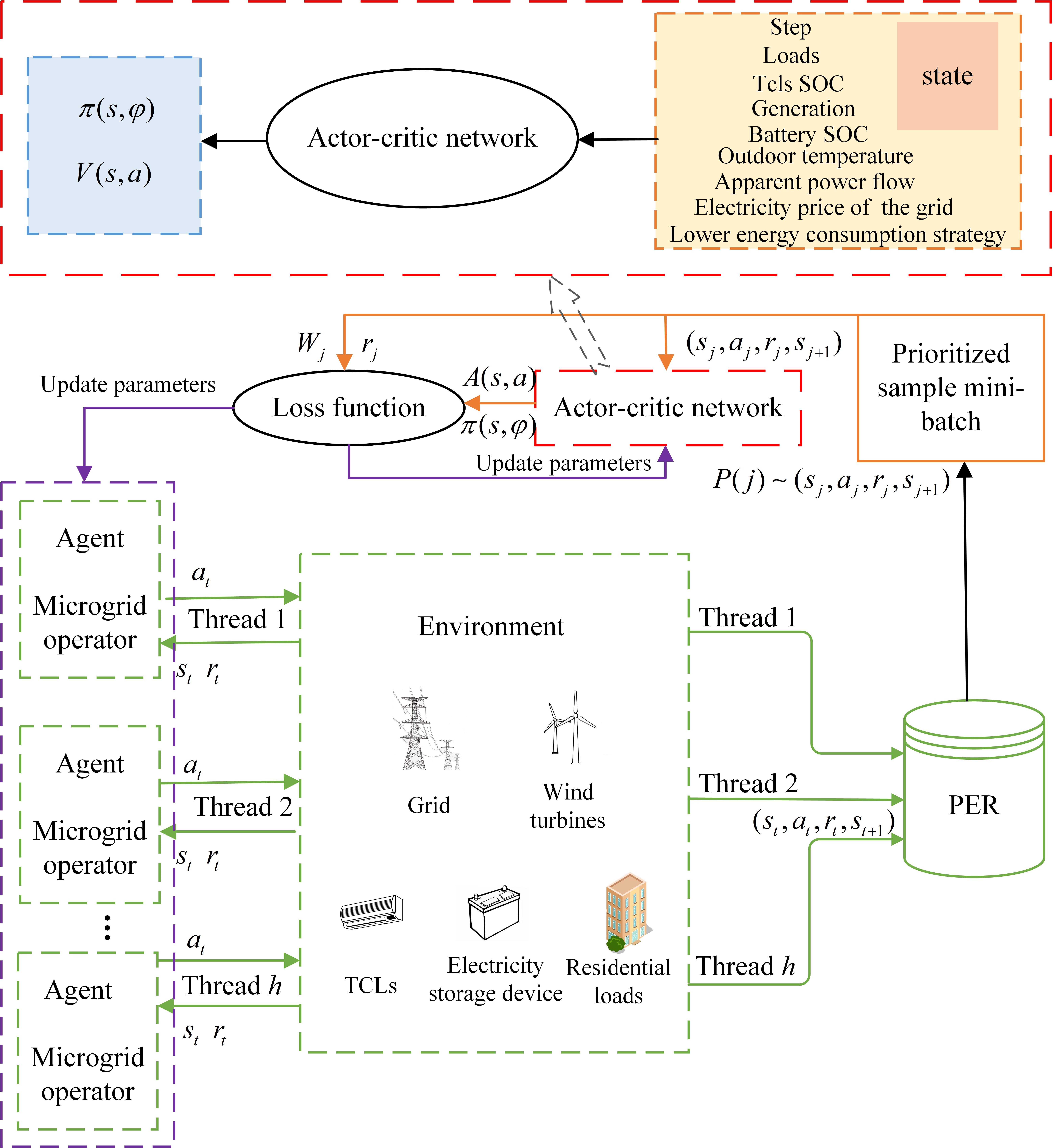}
    \caption{ Workflow of the proposed AutoML-PER-A3C.}
    \label{AutoML-PER-A3C}
\end{figure}

In this section, we introduce the proposed DRL-based method, named AutoML-PER-A3C, and the overall workflow is shown in Fig. \ref{AutoML-PER-A3C}. The key elements related to the application of the AutoML-PER-A3C in the upper level are listed below.

1) Agent: the MG operator is set as an agent. Although in the A3C the agents interact with the environment in a multi-threaded manner, they all refer to the same MG operator.

2) Environment: the environment is composed of TCLs, residential loads, WT, ESD and the main grid.

3) State: state is used to describe the feedback on the environment of actions taken by the agent in the current environment. Specifically, the state consists of the SOC values of TCLs $[SOC_{TCL,t}^1,SOC_{TCL,t}^2,...,SOC_{TCL,t}^n]$, ESD state of charge $SO{C_{ESD,t}}$, grid electricity price ${\lambda _{grid,t}}$ , WT power output ${P_{WT,t}}$, fixed load ${P_{basic,t}}$ in period $t$, time step ${T_{step,t}}$ , outdoor ambient temperature ${T_{out}}(t)$, apparent power flow  ${P_{ab,t}}$ between nodes $a$ and $b$, and the lower-level energy consumption strategy ${P_{TCL,n}}$.

4) Action: action is made of the action space ${A_{loads}}$ of residential price-responsive loads, the action space ${A_{TCLs}}$ of TCLs, the action space  of insufficient energy ${A_{shortage}}$ and the action space   of excess energy ${A_{overplus}}$. In detail,  ${A_{loads}}$ is divided into five actions according to the established price level, namely ${A_{loads}} = \{ a_{loads}^0,a_{loads}^1,a_{loads}^2,a_{loads}^3,a_{loads}^4\}  = \{0, 1, 2, 3, 4\} $. Here, the residential load action corresponds to the price level ${\varsigma _t}$ set by the operator;  ${A_{TCLs}}$ is divided into four actions according to the energy level allocated to TCLs, namely ${A_{TCLs}} = \{a_{TCLs}^0,a_{TCLs}^1,a_{TCLs}^2,a_{TCLs}^3\} = \{0, 1, 2, 3\}$. 
If energy shortage occurs, ${A_{shortage}}$ is divided into two actions according to the priorities that the ESD and main grid provide energy, namely ${A_{shortage}} = \{a_{shortage}^0,a_{shortage}^1\} = \{0,1\}$, If ESD delivers energy first then ${A_{shortage}}$ is 1, if the grid provides energy first, then ${A_{shortage}}$ is 0; otherwise, ${A_{overplus}}$ includes two actions, namely ${A_{overplus}} = \{a_{overplus}^0,a_{overplus}^1\}  = \{0,1\}$. Here, energy shortage denotes that it is required to purchase electricity since WT power outputs cannot meet the load demand, while excess energy refers to that there is excess electricity after the WT power meets the load demand.

5) Reward: the physical informed-inspired reward is set as the profit of the MG operator, as described in (\ref{OF1}). Since the reward is related to the physical decision variable for MG, the reward helps to guide the operator to take better actions.

Moreover, the actor and critic share a neural network with a parameter $\varphi$ to estimate the policy ${\pi _\varphi }$ and state-value function in this work.
In general, the critic part evaluates the policy by minimizing the loss function. The defined loss function is composed of three parts: the loss of the policy ${L_\pi }$ , the loss  of the state-value ${L_V}$ and the regularization term  with policy entropy ${H_{REG}}$. The total loss function is as follows:
{\setlength\abovedisplayskip{2pt}
\setlength\belowdisplayskip{2pt}
\begin{equation}\label{suanfa10}
{L_{TOTAL}}(\varphi ) = {L_\pi }(\varphi ) + {\beta _V}{L_V}(\varphi ) + {\beta _{REG}}{H_{REG}}(\varphi )
\end{equation} }where ${\beta _V}$ and ${\beta _{REG}}$ are the coefficients of the value loss and regularization term with policy entropy, respectively. In terms of policy improvement, we define the objective function $J(\pi )$ as the average return value of the agent in all starting states under  policy $\pi$. According to the policy gradient theory, when the actor is updated, the gradient can be calculated by 
{\setlength\abovedisplayskip{2pt}
\setlength\belowdisplayskip{2pt}
\begin{equation}\label{suanfa11}
{\nabla _\varphi }J(\pi ) = {E_{s \sim {\tau ^\pi },a \sim \pi (s)}}[A(s,a){\nabla _\varphi }\log \pi (a|s)]
\end{equation} }where the advantage function is defined as $A({s_j},{a_j}) = Q({s_j},{a_j}) - V({s_j})$, which describes how much better the action taken in the current state is than a normal situation;  ${\tau ^\pi }$ is the state probability distribution under  policy $\pi $; and $\pi (s)$ is the action probability distribution under  state $s$. To maximize the objective function $J(\pi )$, the definition of the policy loss is
{\setlength\abovedisplayskip{2pt}
\setlength\belowdisplayskip{2pt}
\begin{equation}\label{suanfa12}
{L_\pi }(\varphi ) =  - J(\pi ) =  - \frac{1}{N}\sum\limits_{j = 1}^N {A({s_j},{a_j})\log \pi ({a_j}|{s_j})} 
\end{equation} }

The state-value loss function is given as follows:
{\setlength\abovedisplayskip{2pt}
\setlength\belowdisplayskip{2pt}
\begin{equation}\label{suanfa13}
{L_V}(\varphi ) = \frac{1}{N}\sum\limits_{j = 1}^N {\delta _j^2} 
\end{equation} }
This paper adopts an entropy regularized policy to weigh exploration and utilization, where  ${\beta _{REG}}$ controls the degree of exploration. Therefore, the regularization term is as follows:
{\setlength\abovedisplayskip{2pt}
\setlength\belowdisplayskip{2pt}
\begin{equation}\label{suanfa14}
{H_{REG}}(\varphi ) =  - \frac{1}{N}\sum\limits_{j = 1}^N {g(\pi ({s_j}))}  = \frac{1}{N}\sum\limits_{j = 1}^N {\pi {{(s)}_j}} \log \pi {(s)_j}
\end{equation} }

Utilization of sampled data is also an important topic in DRL research. By storing both historical and contemporary data, the experience replay buffer $\Re $ \cite{mnih2015human} can eliminate the temporal correlation of samples \cite{zhang2017deeper}. However, the traditional experience replay uses uniform sampling, which does not make full use of experience. To solve the problem of low sampling efficiency, this study leverages a PER strategy \cite{schaul2015prioritized}. Specifically, the priority of each sample is proportional to the absolute value of temporal difference
errors (TD-errors). In the prioritized sampling of a minibatch of $N$ transitions $\left\{ {({s_j},{a_j},{r_j},{s_{j + 1}})} \right\}_j^N,j = 1,2,...N$, we define the probability of each sample $j$ being sampled as follows:
{\setlength\abovedisplayskip{2pt}
\setlength\belowdisplayskip{2pt}
\begin{equation}\label{suanfa15}
P(j) = \frac{{p_j^\upsilon }}{{\sum\nolimits_k {p_k^\upsilon } }}
\end{equation} }where ${p_j} = |{\delta _j}| + \xi $, $\xi $ is a very small constant; $|{\delta _j}|$ is absolute TD-error; $\nu $ is used to control the preference of sampling. To correct deviations, we introduce important sampling weights and annealing factors. The weights are set as 
{\setlength\abovedisplayskip{2pt}
\setlength\belowdisplayskip{2pt}
\begin{equation}\label{suanfa17}
{\omega _j} = \frac{{{{({S_\Re }P(j))}^{ - \phi }}}}{{{{\max }_k}({\omega _k})}} = {\left( {\frac{{{{\min }_k}P(k)}}{{{P_j}}}} \right)^\phi }
\end{equation} }where ${S_\Re }$ is the size of the $\Re $, annealing factor $\phi $ is used to correct the flexibility of the amount, and ${\max _k}({\omega _k})$ is to standardize sampling weights. Based on the above analysis, the final AutoML-PER-A3C is summarized in Algorithm \ref{alg:code}.

\begin{algorithm}[!t]\label{code} % 
\small
\renewcommand{\algorithmicrequire}{\textbf{Initialize:}}
\renewcommand{\algorithmicensure}{\textbf{return}}
    \caption{AutoML-PER-A3C Algorithm for Each Thread} 
    \label{alg:code}
	\begin{algorithmic}[1] 
	\REQUIRE the actor-critic neural network parameters $\varphi $.
	\REQUIRE the replay buffer $\Re $ with size ${S_\Re } $.
    \FOR{trial = 1, ..., $M$}
    \STATE Select a set of hyperparameters from the search space according to the Metis Tuner.
    \FOR{episode=1, ..., $E$} 
    \STATE Select random action ${a_t}$ from the action space.
    \STATE Select initial state ${s_t}$ from state space.
    \FOR{$t$=1, ..., $H$}
    \STATE Select action ${a_t}$ from the action space according to the $\varepsilon-greedy$ policy.
    \STATE The lower-level determines the energy consumption strategy used as the new state $s{}_{t + 1}$ according to the price issued by the operator. The agent observes  new state $s{}_{t + 1}$ and then reward ${r_t}$ is calculated using (\ref{OF1}).
    \STATE Store transition $({s_t},{a_t},{r_t},{s_{t + 1}})$ in replay buffer $\Re $ with maximal priority ${P_t} = {\max _{k < t}}{P_k}$.
    \FOR{$j$=1, ..., $N$}
    \STATE Sample transition with probability $P(j)$ using  (\ref{suanfa15}).
    \STATE Compute importance-sampling weight ${\omega _j}$ using  (\ref{suanfa17}) and TD-error ${\delta _j}$.
    \STATE Update the priority of transition according to absolute TD-error $|{\delta _j}|$ .
    \ENDFOR
    \STATE Update actor-critic network according to minimize the loss function, and then pass the parameters of the global network to the worker. 
    \ENDFOR
    \ENDFOR
    \STATE Collect the average reward and upload it to the Metis Tuner.
    \ENDFOR
    \STATE Select the best hyperparameters and policies according to the maximum average reward.
  \end{algorithmic}
\end{algorithm}

\subsection{Solution Process}   %
The solution process of the MG dispatching model is described in detail as follows:

Step 1: According to (\ref{OF1})-(\ref{Network4}), construct the optimal scheduling model of the upper-level operator;

Step 2: Set and update the episode of AutoML-PER-A3C training;

Step 3: Execute the markov decision process in the DRL;

Step 4: Obtain the optimal strategy of the upper-level operator, and release the price to the lower level; %

Step 5: Build an optimal scheduling model of the lower level
according to  (\ref{xiacengmubiao})-(\ref{llyueshu9});

Step 6: Enter the MG parameters;

Step 7: Use the DOCPLEX optimizer to solve the lower-level scheduling model;

Step 8: Determine whether the solution exists. If it exists, the users’ energy consumption is passed to the upper level as the new state of the DRL; otherwise, return to Step 6; 

Step 9: Determine whether the termination condition is met. If met, proceed to execute; otherwise, return to Step 2;

Step 10: Obtain the optimal MG scheduling scheme.

\section{Case Study}\label{Case Study}
\begin{figure}[htbp]
\begin{minipage}[t]{0.49\linewidth}
\centering
\includegraphics[height=3.0cm,width=3.9cm]{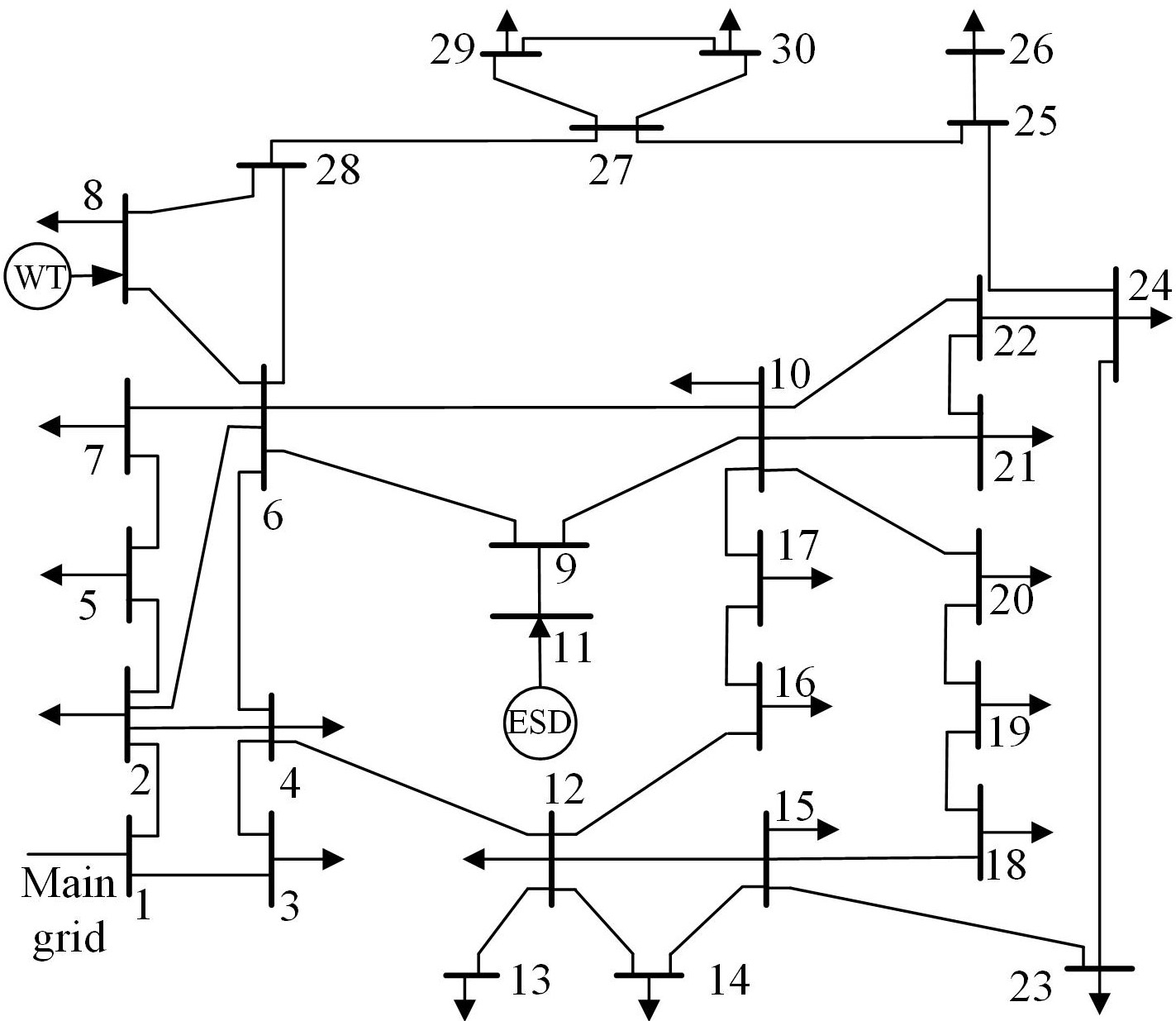}
\caption{MG one-line graph using modified IEEE 30-bus system.}
\label{IEEE30}
\end{minipage}%
\hfill
\begin{minipage}[t]{0.48\linewidth}
\centering
\includegraphics[height=3.0cm,width=3.9cm]{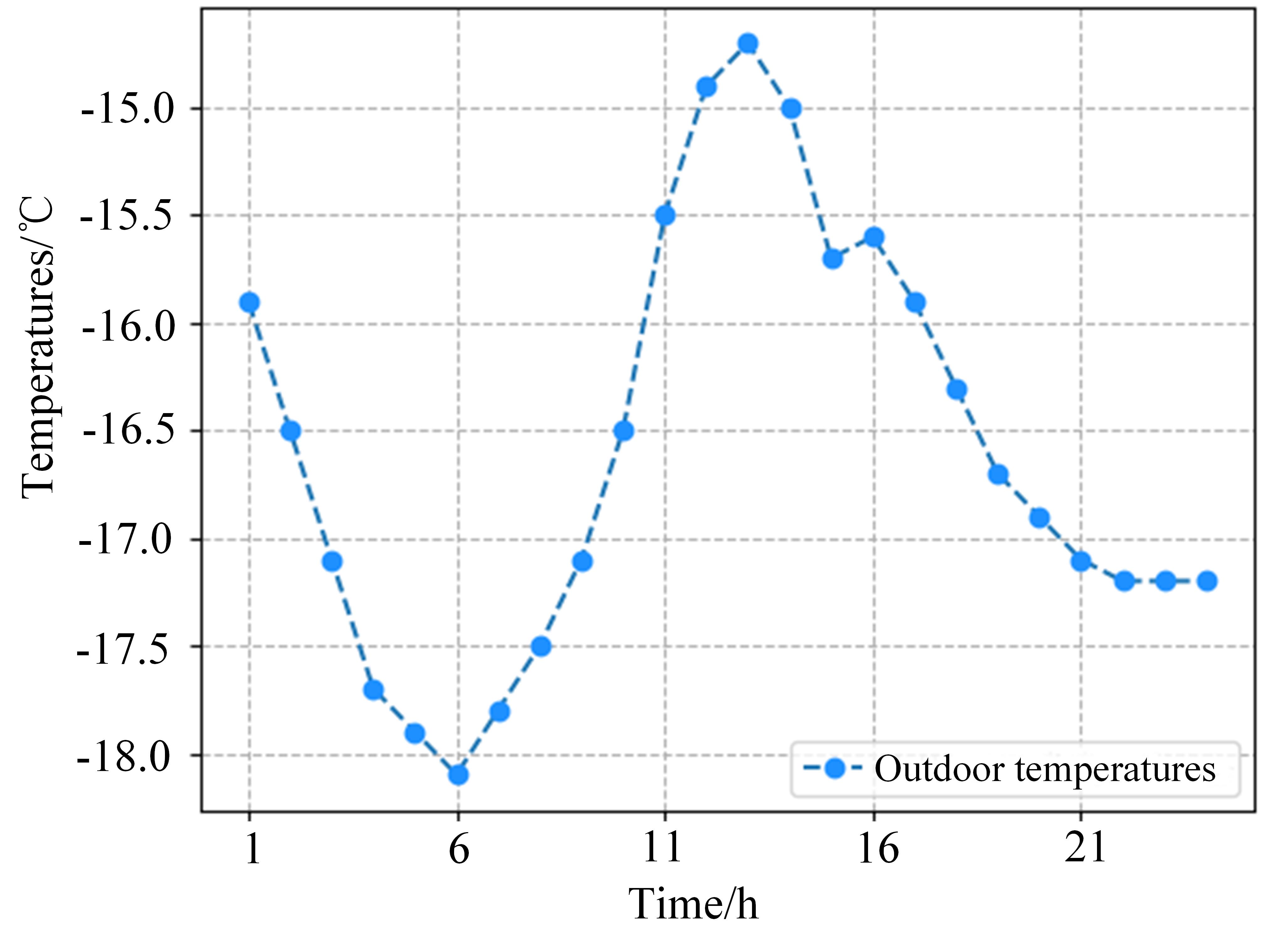}
\caption{{Change of the outdoor temperatures.}}
\label{wenduquxian}
\end{minipage}
\end{figure}

To verify the effectiveness of the proposed scheduling model and method, this paper uses the improved IEEE 30-bus test system to test on the microgrid, as shown in Fig. \ref{IEEE30}. Moreover, the proposed AutoML-PER-A3C algorithm used in the upper level has been implemented using Tensorflow 1.10 and Keras 2.23 in python 3.6, and the lower-level problem has been addressed using the DOCPLEX optimizer python interface. All simulation tests are carried out on a PC platform with Intel Core i5-6300HQ CPU (2.3 GHz) and 8GB RAM.

\subsection{Settings in Test Case }
In this paper, we consider a test case of a MG, whose key components include 150 residential loads, 100 TCLs, and an ESD. Table \ref{tab3} describes the main parameter settings of the MG. Moreover, the data record of WT powers is provided by Fortum Oyj of a wind farm in Finland. The price data of purchasing and selling electricity comes from the FINGRID database. Market price ${\lambda _{market}}$ is taken as 5.53 € Cents/kWh, which is from the Helsinki market in January 2018. Fig. \ref{wenduquxian} shows the daily outdoor temperature curve.

\begin{table}[htbp]
  \centering
  \caption{Setting of the main microgrid parameters}\label{tab3}
  \begin{tabular}{cccc}
    \toprule
    Parameters & Values & Parameters & Values \\
    \midrule
    $C_{air,n}$                & $\mathcal{N}(0.004,0.0008)$ & $T_{UB,n} (^{\circ}C)$   & 25 \\
    $C_{m,n}$                  & $\mathcal{N}(0.3,0.004)$    & $\eta_{ch}, \eta_{dc}$   & 0.9 \\
    $P_{TCL,n}$ (kW)           & $\mathcal{N}(1.5,0.01)$     & $P_{ch,\max}, P_{dc,\max}$ (kW) & 250 \\
    $Q_{air}$                  & $\mathcal{N}(0.0,0.01)$     & $C_{ESD,\max}$ (kWh)     & 500 \\
    $\eta$                     & $\mathcal{N}(10,6)$         & $\lambda_{gen}$ (€/MW)   & 32 \\
    $\sigma_t$                 & $\mathcal{N}(0.4,0.3)$      & $\mu_{import}$ (€/MW)    & 9.7 \\
    $\varsigma_t$              & $\{-2, -1, 0, 1, 2\}$       & $\mu_{export}$ (€/MW)    & 0.9 \\
    $\iota$                    & $\{0, 40, 80, 120\}$        & $T_{LB,n} (^{\circ}C)$   & 19 \\
    \bottomrule
  \end{tabular}
\end{table}%

\subsection{Results and Analysis}
\subsubsection{Analysis of Optimization Results Using the AutoML}
\begin{figure}[t]
    \centering
    \includegraphics[width=3.3in,height=0.8in]{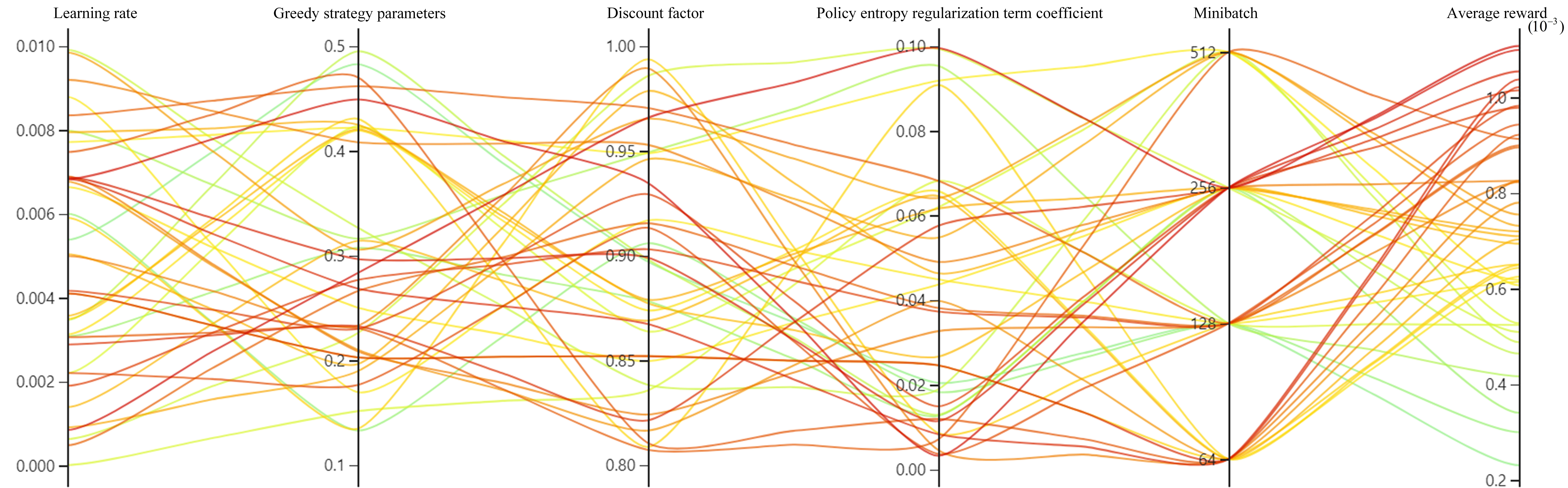}
    \caption{Hyperparameter optimization results using the AutoML.}
    \label{Automlsuanli}
\end{figure}

Simulation experiments were carried out to assess the efficacy of the AutoML. The existing hyperparameter determination's intermediate outcomes are evaluated by the AutoML, which then makes logical recommendations for the subsequent hyperparameter trial. Finally, Fig. \ref{Automlsuanli} shows the WebUI's hyperparameter being selected outcomes for each trial.

Fig. \ref{Automlsuanli}  displays the outcomes of AutoML optimization for the PER-A3C required hyperparameters. The band of values of each hyperparameter is shown as an ordinate, and the average reward value obtained by applying these hyperparameters is shown as the final ordinate. Additionally, a deeper shade of red indicates that the given hyperparameters are optimal. The final assessment's average reward shows that the AutoML we developed is effective in optimizing problems. This illustrates that AutoML is successful in identifying optimal hyperparameter combinations for A3C, which enhances A3C's generalization capability and learning quality.

\subsubsection{Demand Response Analysis}

\begin{figure}[htbp]
\begin{minipage}[t]{0.49\linewidth}
\centering
\includegraphics[height=3.3cm,width=4.2cm]{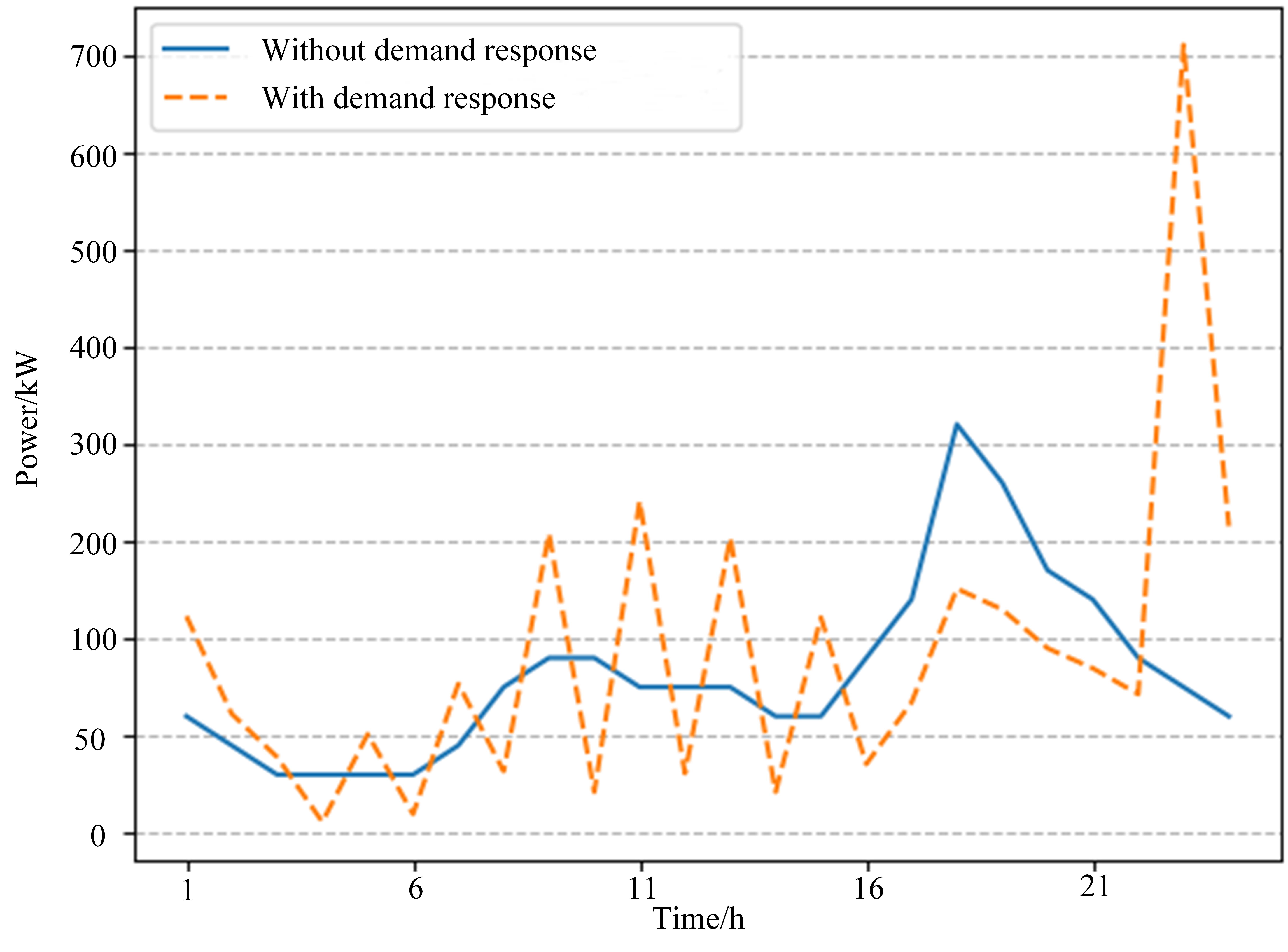}
\caption{Changes of residential load  with and without demand response.}
\label{residential electrical load}
\end{minipage}%
\hfill
\begin{minipage}[t]{0.48\linewidth}
\centering
\includegraphics[height=3.3cm,width=4.2cm]{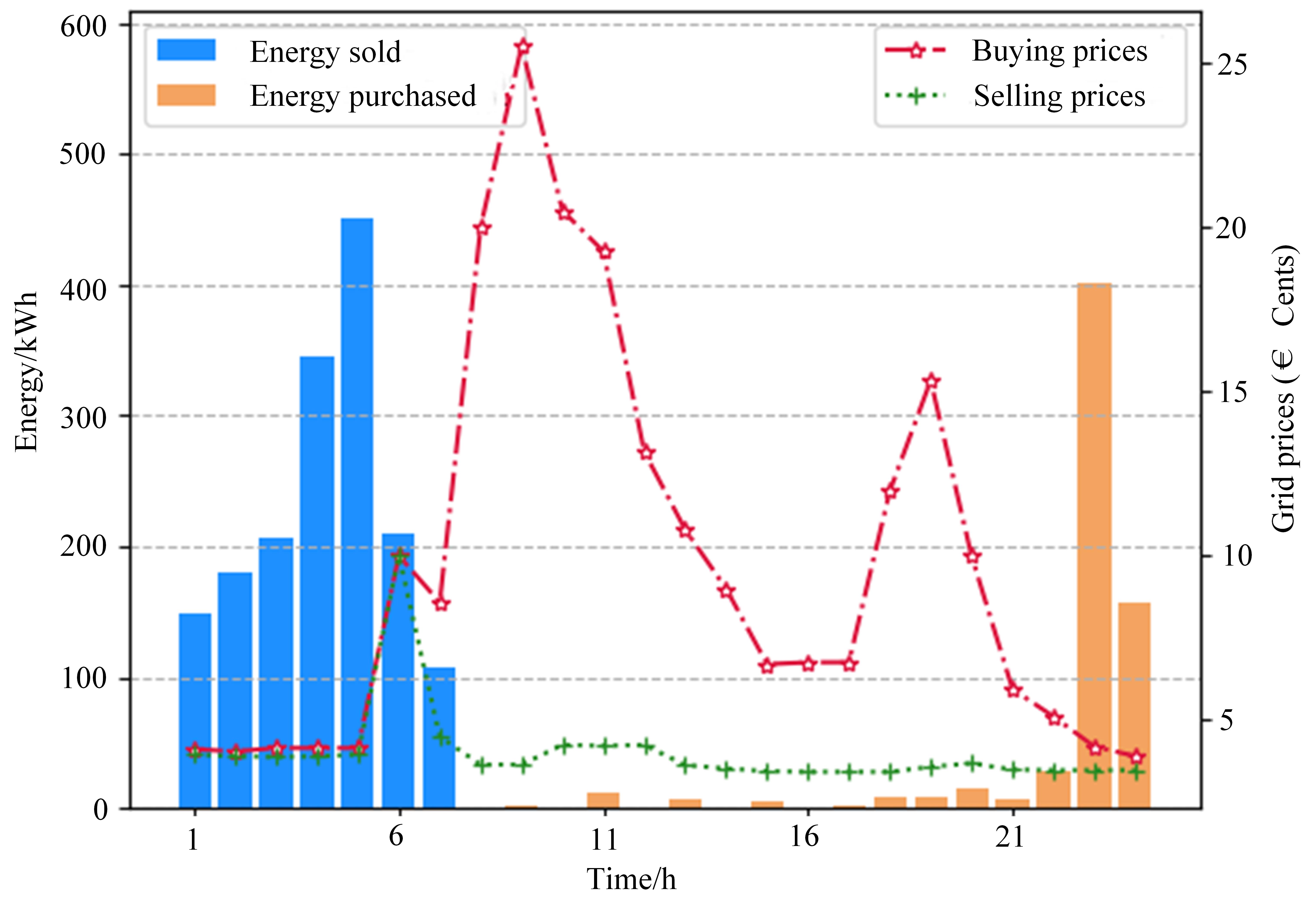}
\caption{Transaction electricity and  prices with the grid.}
\label{energy exchanged}
\end{minipage}
\end{figure}

To test the effect of demand response, simulation analysis has been carried out with the results shown in  Fig. \ref{residential electrical load}, which indicates that the residential load decreases during peak hours and increases during off-peak hours, reflecting the active participation of residential customers in demand response. Therefore, by leveraging price signals to guide resident  behaviors, demand response manages to achieve peak-shaving effects.

As shown in Fig. \ref{energy exchanged}, when the operator trades with the main grid, transaction electricity is arranged according to the fluctuation of electricity prices. For example, the electricity sold in period 1:00-7:00 and electricity purchased in period 22:00-24:00 are more than other periods. The main reasons for this phenomenon are summarized as follows. (1) During off-peaking hours, WT generate more electricity, and the operator sells the excess electricity to the grid for more profits. (2) During peaking hours, residents participate in demand response by shifting the load, thereby the operator purchases less electricity to meet the load demands.  In this way, the scheduling strategy obtained by the AutoML-PER-A3C successfully balances the interests of residents and operator.

\subsubsection{Energy Allocation and Consumption Analysis of TCLs}
\begin{figure}[htbp]
\begin{minipage}[t]{0.49\linewidth}
\centering
\includegraphics[height=3.4cm,width=4.3cm]{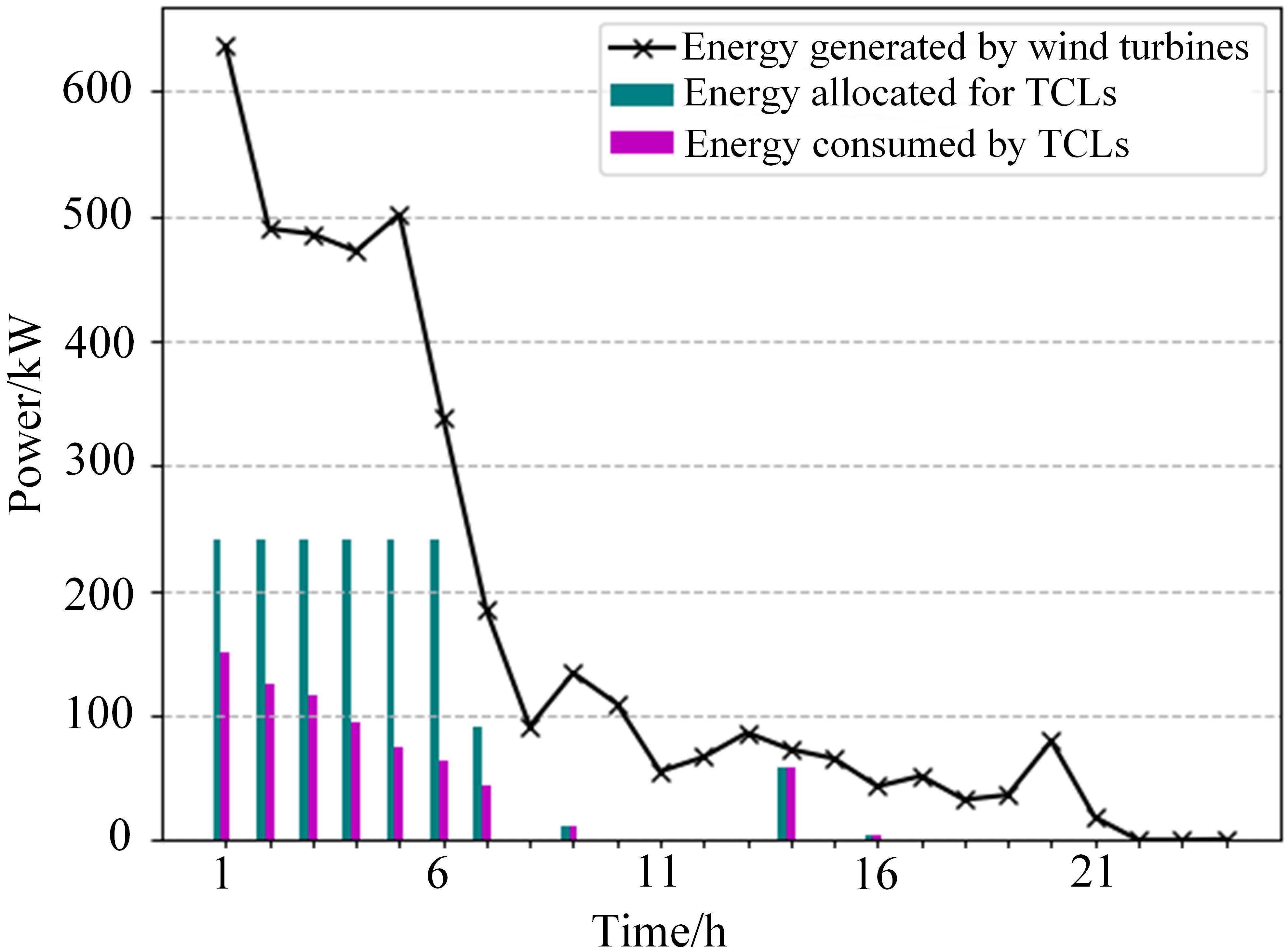}
\caption{Energy allocated for TCLs and energy consumed by TCLs.}
\label{tclsuanli}
\end{minipage}%
\hfill
\begin{minipage}[t]{0.48\linewidth}
\centering
\includegraphics[height=3.4cm,width=4.3cm]{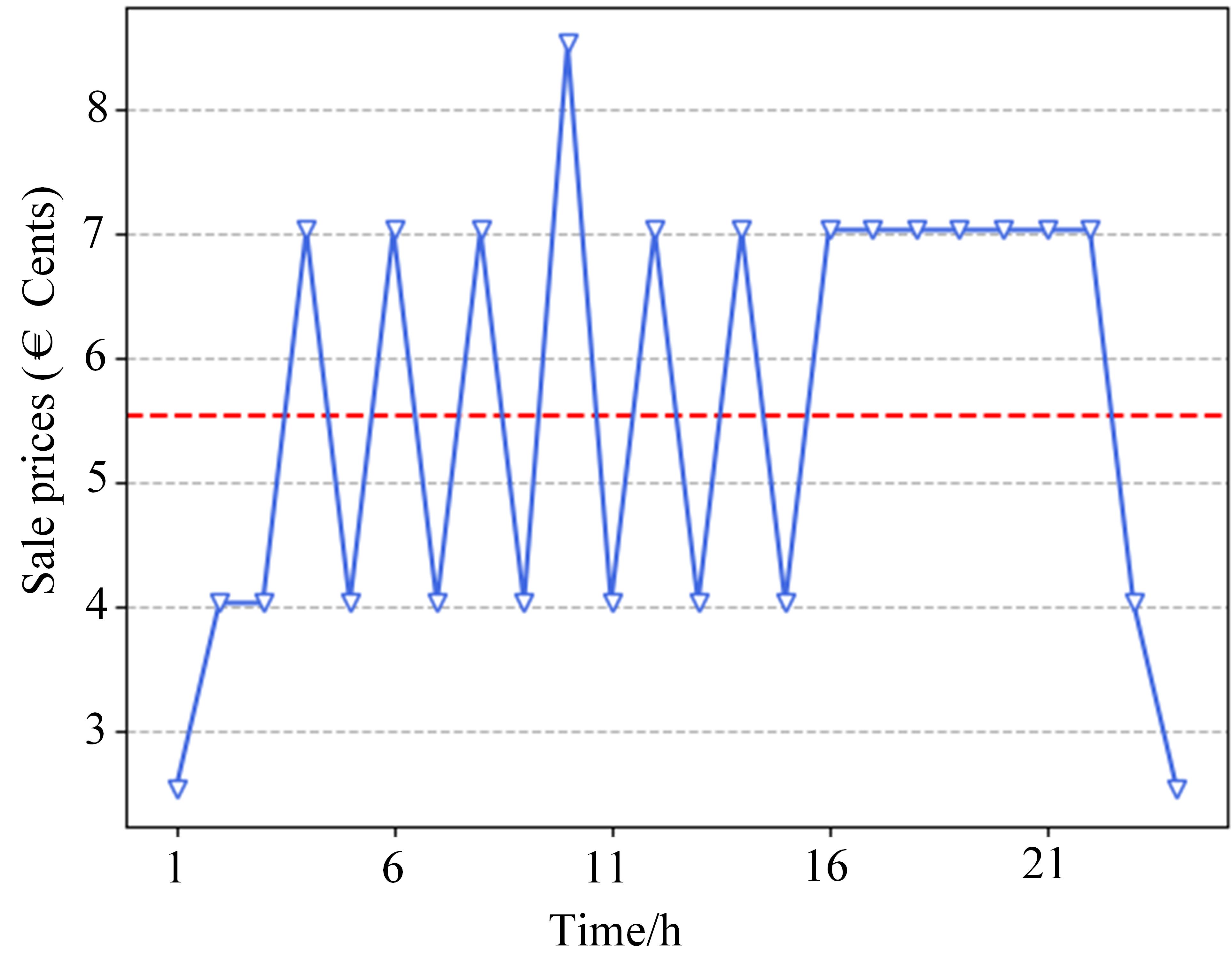}
\caption{Change of the prices formulated by the operator.}
\label{Price strategy}
\end{minipage}
\end{figure}

For purpose of examining the thermal flexibility of TCLs, their energy allocation and consumption strategies have been investigated. Fig. \ref{tclsuanli} shows the policies of energy allocated for TCLs and energy consumed by TCLs, which are assigned to TCLs by the AutoML-PER-A3C. The energy allocated for TCLs is set within a reasonable range, whose lower bound meets the demand of the load. In each time period, the DRL selects the policy of the corresponding  TCLs' level according to the action. It can be seen that to maintain a comfortable temperature in period 13:00-24:00 without  needing to provide more energy, the DRL manages to allocate more energy in period 1:00-12:00. The above shows that the DRL is able to significantly improve energy utilization efficiency by making full use of the thermal flexibility of TCLs.

\subsubsection{Analysis of the Pricing Mechanism} %
To analyze the effectiveness of the pricing mechanism used, the mechanism has been studied. Fig. \ref{Price strategy} shows the change of the prices formulated by the operator, where the red dotted line denotes the benchmark market price. The operator improves electricity prices during peaking hours and reduces the prices during off-peaking hours, which prevents  residential users purchasing large amounts of electricity in peaking hours and thereby reduces the electricity cost of residents. Therefore, the pricing mechanism obtained by the DRL is capable of trading off the interests of residential users and operator.

To further examine the superiority of our pricing mechanism, this section performs the comparison of three  different pricing mechanisms.

1) Mechanism 1 (our mechanism): the pricing mechanism can be obtained by solving the bi-level model.

2) Mechanism 2 (time-of-use price): the time-of-use (TOU) price is divided into day-time and night-time prices. For users, there are only these two types of electricity prices available.

3) Mechanism 3 (market price): the electricity price is a fixed 24-hour price ${\lambda _{market}}$.

Table \ref{tab4} shows the comparison results of different pricing mechanisms. Ones can see that comparing mechanisms 2 and 3,   the 10-day average profit of the  operator has respectively increased by 25.3$\%$ and 39.4$\%$ in the proposed pricing mechanism. As a result, it can be  drawn that the our pricing mechanism can ensure the economy of MG operations.

\begin{table}[htbp]
  \centering
  \caption{Comparison of the average profits of different pricing mechanisms}\label{tab4}
    \begin{tabular}{l|c}   % 
    \hline
    \multicolumn{1}{l|}{\textbf{Pricing mechanism}} & \textbf{Profit (€)} \\
    \hline
    \textbf{Mechanism 1:} Our pricing mechanism & 1050.00 \\
    \hline
    \textbf{Mechanism 2:} TOU price & 838.00 \\
    \hline
    \textbf{Mechanism 3:} Market pricing  & 753.00 \\
    \hline
    \end{tabular}%
\end{table}%

Fig. \ref{different pricing} shows the daily profits of the operator in the three pricing mechanisms. As illustrated in this figure, mechanism 1 obviously outperforms the other  mechanisms in terms of the profits. Therefore, our pricing mechanism is better than the other mechanisms.

\subsubsection{Analysis of WT Powers  and Transaction  Prices with  Main Grid}
\begin{figure}[htbp]
\begin{minipage}[t]{0.49\linewidth}
\centering
\includegraphics[height=3.5cm,width=4.4cm]{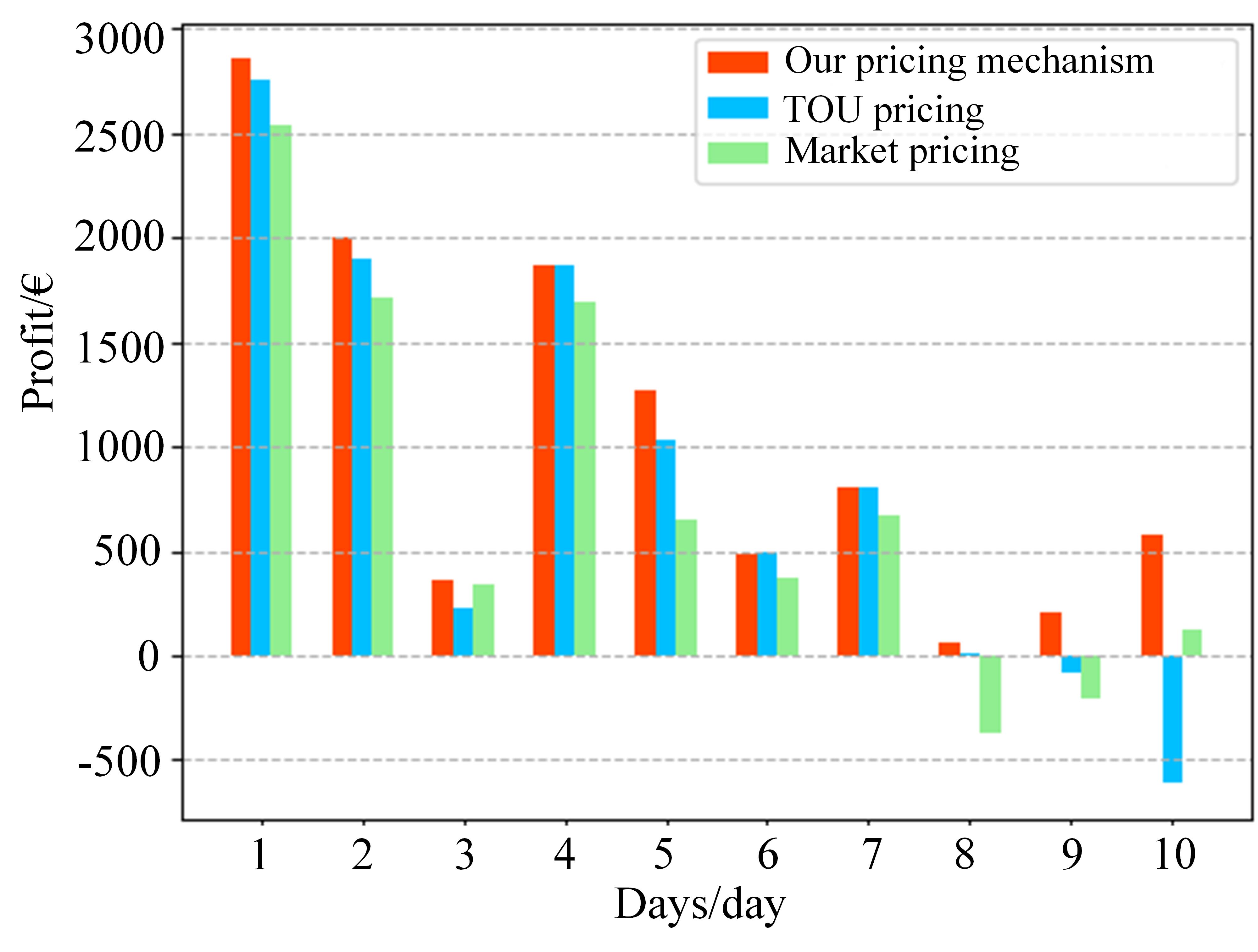}
\caption{Comparison of daily revenue under different pricing mechanisms.}
\label{different pricing}
\end{minipage}%
\hfill
\begin{minipage}[t]{0.48\linewidth}
\centering
\includegraphics[height=3.5cm,width=4.4cm]{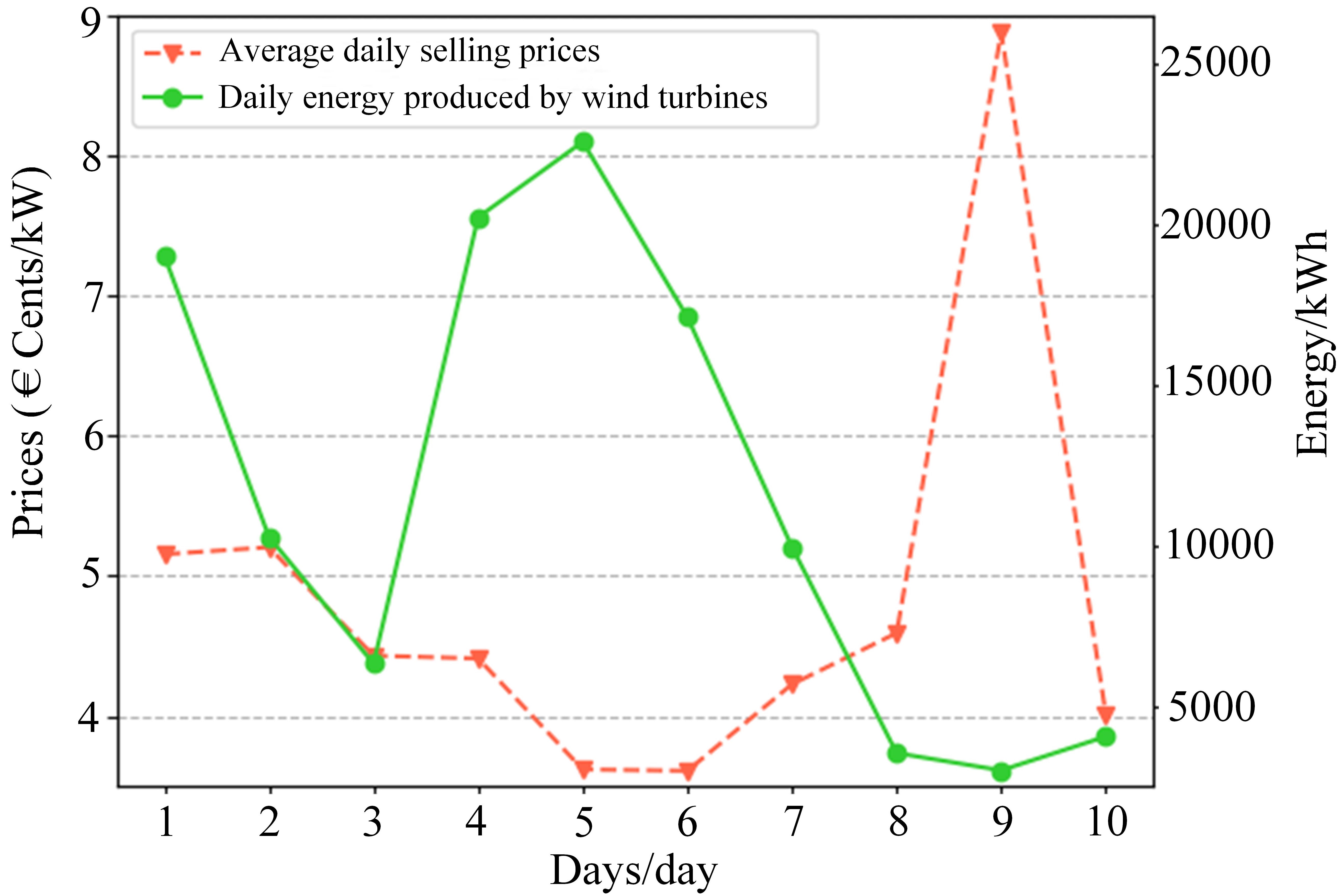}
\caption{WT power and transaction prices with the main grid.}
\label{fanbi}
\end{minipage}
\end{figure}

To properly evaluate the impact of renewable generation on  electricity market, the  relationship between them is studied by simulation analysis with the results shown in Fig. \ref{fanbi}. The figure suggests that the power outputs of WT are negatively correlated with the average daily electricity selling price.

\subsubsection{ Analysis of Charging-Discharging Strategies of the ESD}
\begin{figure}[htbp]
\begin{minipage}[t]{0.49\linewidth}
\centering
\includegraphics[height=3.5cm,width=4.4cm]{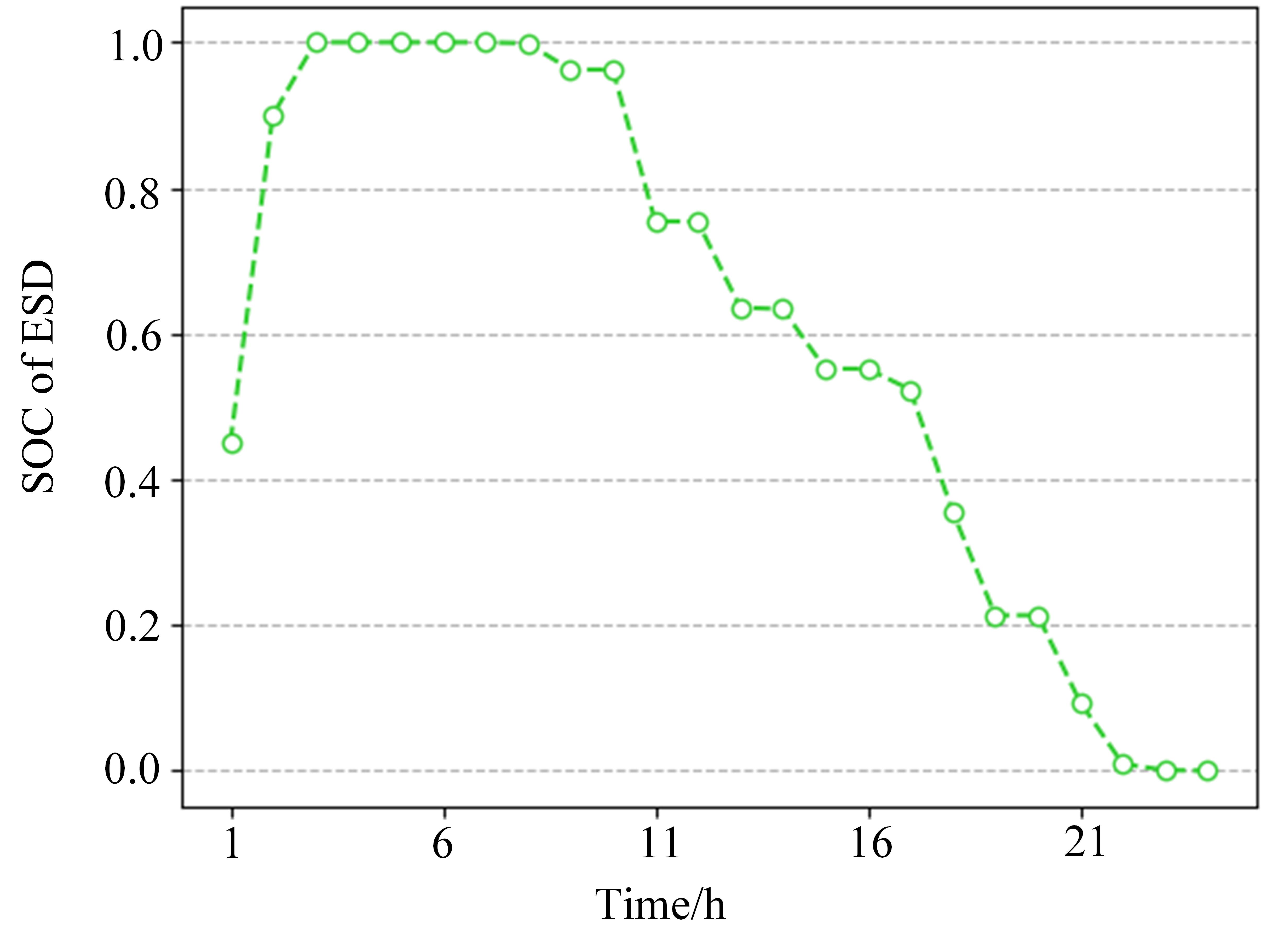}
\caption{Change of charging and discharging states of the ESD.}
\label{ESDsuanli}
\end{minipage}%
\hfill
\begin{minipage}[t]{0.48\linewidth}
\centering
\includegraphics[height=3.5cm,width=4.4cm]{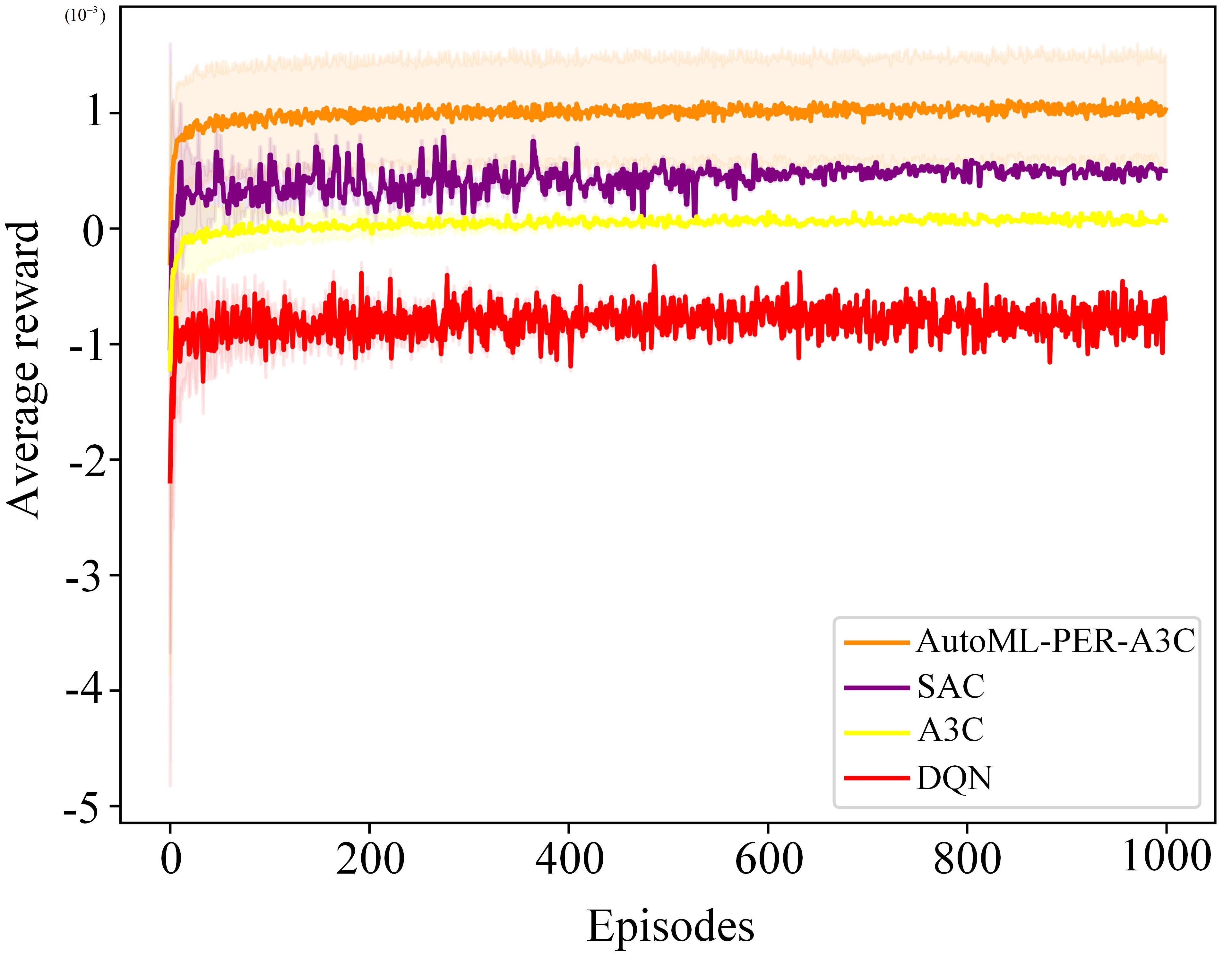}
\caption{Comparison of average reward values of different RL algorithms.}
\label{RLduibi}
\end{minipage}
\end{figure}

To verify the effectiveness of the charging and discharging strategies of the ESD, the simulation analysis has been performed and the results are shown in Fig. \ref{ESDsuanli}. 

From this figure, ones can observe that the ESD stores amounts of energy in non-peaking hours, and releases the energy during the peaking hours. In the initial stage, WT generate sufficient power and the ESD is quickly charged; and then, the ESD's energy is gradually reduced to meet the growing load demand. The above analysis shows that the charging and discharging strategies of the ESD fully consider the energy shortage in future peaking hours, thereby improving the flexibility of the system operation.

\subsubsection{Performance Comparison of RL Methods}
To demonstrate the superiority of the proposed DRL, comparison tests with other RL methods have been carried out. Fig. \ref{RLduibi} shows the 10-day operator's average profit (i.e., average reward) obtained by applying each RL method to the built upper-level model. In the first learning step for each RL approach, the agent randomly explores numerous paths, which does not always result in behaviors that are more profitable, hence the average profit is negative. And then profits for all RL strategies turn positive and keep rising as more data is collected, finally converging at some stage.
Therefore, the accumulated physical informed-inspired reward values help the agent learn better policies during training, that is, the reward can indirectly explain the rationality of the agent's policies.

Regarding the average profit of the operator, the proposed method is significantly superior to other RL alternatives. In Fig.  \ref{RLduibi}, the average profit of the proposed method increases by 81.0$\%$ compared to soft actor-critic (SAC). And the average profit value of original A3C is close to 0, which is far inferior to the performance of the proposed method and SAC. Moreover, the average profit value of deep Q network (DQN) is negative, which performance is not as good as the above algorithms. The main reasons are two-fold: (1) in this paper, the PER improves generalization performance of the original A3C; (2) the AutoML  avoids the deviation caused by human experience exploration to adjust the hyperparameters. Therefore, it can be seen that our method has higher economy than alternative RL approaches, and that the policy improvement of the original A3C has significant advantages.

As traditional single-thread RL is generally difficult to adapt to massive and intractable tasks, it is critical for RL to speed up the learning process \cite{li2023deep2}. For this reason, the calculation efficiencies of the RL methods are studied with the results listed in Table \ref{tab5}. For the average training time of each episode, the training time of the AutoML-PER-A3C is significantly less than other RL methods due to the multi-thread training and efficient sampling of the PER. Therefore, due to its superior computational efficiency, the proposed approach surpasses other  alternatives.

\begin{table}[htbp]
  \centering
  \caption{Comparison of calculation performance of RL methods}
  \label{tab5}
  \setlength{\tabcolsep}{2pt} % 进一步减少列间距
  \begin{tabular}{cccc}
    \toprule
    \textbf{Method} & 
    \begin{tabular}[c]{@{}c@{}}\textbf{Average training} \\ \textbf{time per} \\ \textbf{episode (min)}\end{tabular} & 
    \begin{tabular}[c]{@{}c@{}}\textbf{Number of} \\ \textbf{episodes}\end{tabular} & 
    \begin{tabular}[c]{@{}c@{}}\textbf{Total training} \\ \textbf{time (min)}\end{tabular} \\
    \midrule
    \textbf{DQN}   & 47.20 & 16 & 755.20 \\
    \midrule
    \textbf{SAC}   & 45.21 & 16 & 723.36 \\
    \midrule
    \textbf{A3C}   & 0.726 & 16 & 11.62 \\
    \midrule
    \textbf{AutoML-PER-A3C} & 0.724 & 16 & 11.58 \\
    \bottomrule
  \end{tabular}
\end{table}

\subsubsection{Analysis of optimization results with and without one-hot encoding}
\begin{figure}[t]
    \centering
    \includegraphics[width=2.3in,height=1.6in]{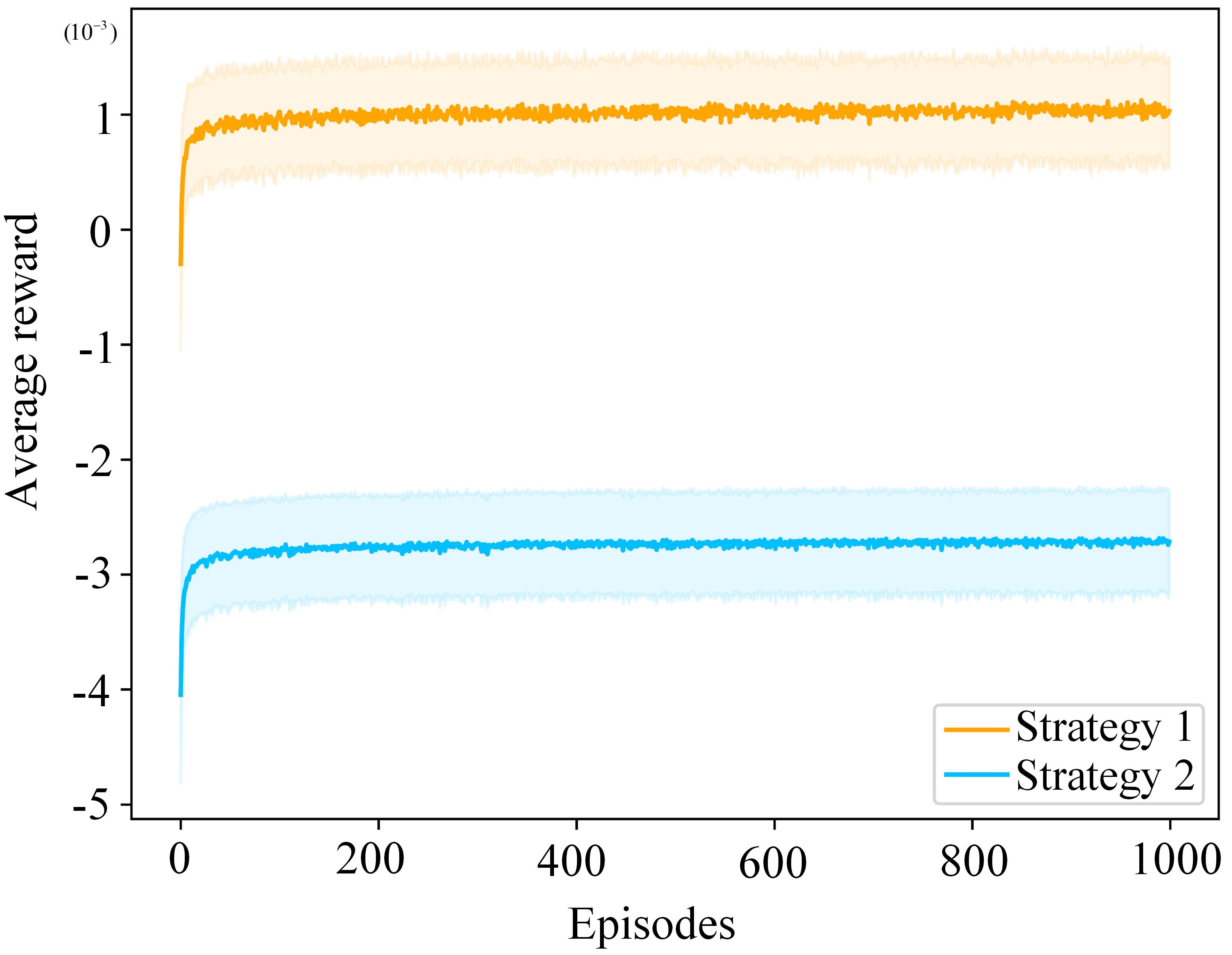}
    \caption{Analysis of optimization results with and without one-hot encoding.}
    \label{onehot}
\end{figure}

To verify the effect of one-hot encoding on the performance of the DRL algorithm, the following simulation analysis is performed. Fig.\ref{onehot} demonstrates the impact on the DRL learning process with and without one-hot encoding, where one-hot encoding is applied to the DRL in strategy 1, however, it is not applied in strategy 2. From the figure, it can be seen that in the initial stage of learning, the reward values are all negative due to the random exploration of the agent. As the experience accumulates, strategy 1 tends to converge to a positive value, while strategy 2 is converging to a negative value. Therefore, one-hot encoding is more in line with the perception of neural networks, which is more favorable to the learning and training process of DRL and gets higher reward values. The one-hot encoding significantly improves the quality of learning.

\section{Conclusion}\label{Conclusion}
To coordinate the interests of operator and users in a MG under complex, changeable and uncertain operating conditions, this work proposes a new MG scheduling model considering the thermal flexibility of thermostatically controlled loads and demand response by leveraging a reinforcement learning based bi-level programming. Based on the simulation results, the following conclusions can be safely drawn:

1) The proposed bi-level MG scheduling model manages to balance the interests of multiple stakeholders through  leveraging the thermal flexibility of TCLs and demand response.

2) The developed deep reinforcement learning theory-based optimization solution method combining AutoML-PER-A3C and DOCPLEX is proven to be effective for addressing the proposed bi-level programming problems. Furthermore, the reinforcement learning enables our approach to be particularly suitable for handling complex and diverse tasks in MG.  

3) The study's findings confirm that the suggested method outperforms other contemporary reinforcement learning options in terms of both economic viability and computational efficiency. And the proposed method can achieve good convergence effectiveness by repeated learning.

\section*{Appendix A}\label{secA1}
\section*{ESD Model}\label{secA2}
At time $t+1$, the ESD's charge-discharge power is proportional to its available capacity, as shown in the following expression:
{\setlength\abovedisplayskip{2pt}
\setlength\belowdisplayskip{2pt}
\begin{equation}\label{ESD1}
{C_{ESD,t + 1}} = {C_{ESD,t}} + ({\eta _{ch}}{P_{ch,t}} - {P_{dc,t}}/{\eta _{dc}})\Delta t{\rm{  }} \quad \forall t
\end{equation} }where ${\eta_{ch}}$ and ${\eta_{dc}}$ denote the charging and discharging rates of the ESD; ${P_{ch,t}}$ and ${P_{dc,t}}$ are the charging and discharging powers for ESD at time $t$; ${C_{ESD,t}}$ and ${C_{ESD,t + 1}}$ represent the available capacity of the ESD at time $t$ and $t+1$, respectively. Moreover, for on-line ESD power monitoring, it is further determined that the state of charge $SO{C_{ESD,t}}$ of the ESD in period $t$ is as follows.

{\setlength\abovedisplayskip{2pt}
\setlength\belowdisplayskip{2pt}
\begin{equation}\label{ESDSOC}
SO{C_{ESD,t}} = {{{C_{ESD,t}}} \mathord{\left/
 {\vphantom {{{C_{ESD,t}}} {{C_{ESD,\max }}}}} \right.
 \kern-\nulldelimiterspace} {{C_{ESD,\max }}}} \quad \forall t
\end{equation} }where ${C_{ESD,\max }}$ is the ESD’s  maximum carrying capacity.

\section*{Appendix B}\label{secB1}
\section*{ESD Constraints}\label{secB2}
Charge and discharge power limits: allowed charging and discharging powers of the ESD are capped at safe levels. The power constraints are as follows:
{\setlength\abovedisplayskip{2pt}
\setlength\belowdisplayskip{2pt}
\begin{equation}\label{ESDyueshu1}
\left\{ \begin{array}{l}
0 \le {P_{dc,t}} \le {P_{dc,\max }}\\
0 \le {P_{ch,t}} \le {P_{ch,\max }}
\end{array} \right.{\rm{  }}\forall t
\end{equation} }%

Capacity limits: the ESD capacity is restricted to a specific area in order to prolong battery life.
{\setlength\abovedisplayskip{2pt}
\setlength\belowdisplayskip{2pt}
\begin{equation}\label{llyueshu5}
{C_{ESD,\min }} \le {C_{ESD,t}} \le {C_{ESD,\max }}{\rm{  }}\quad \forall t
\end{equation} }

Starting and ending limits: to guarantee that each dispatching cycle begins with the same parameters, the ESD must adhere to the following bounds:
{\setlength\abovedisplayskip{2pt}
\setlength\belowdisplayskip{2pt}
\begin{equation}\label{llyueshu6}
{C_0} = {C_{{T_{end}}}} = {C_{ESD,\min }}
\end{equation} }where ${C_0} = 0$ and ${C_{{T_{end}}}}$ represent the ESD's available storage space at the start and finish of a given dispatching round $T$ (In this study, $T$ has been defined as 24 hours).

\bibliographystyle{IEEEtran}
\bibliography{publication.bbl}

\begin{thebibliography}{10}
\providecommand{\url}[1]{#1}
\csname url@samestyle\endcsname
\providecommand{\newblock}{\relax}
\providecommand{\bibinfo}[2]{#2}
\providecommand{\BIBentrySTDinterwordspacing}{\spaceskip=0pt\relax}
\providecommand{\BIBentryALTinterwordstretchfactor}{4}
\providecommand{\BIBentryALTinterwordspacing}{\spaceskip=\fontdimen2\font plus
\BIBentryALTinterwordstretchfactor\fontdimen3\font minus
  \fontdimen4\font\relax}
\providecommand{\BIBforeignlanguage}[2]{{%
\expandafter\ifx\csname l@#1\endcsname\relax
\typeout{** WARNING: IEEEtran.bst: No hyphenation pattern has been}%
\typeout{** loaded for the language `#1'. Using the pattern for}%
\typeout{** the default language instead.}%
\else
\language=\csname l@#1\endcsname
\fi
#2}}
\providecommand{\BIBdecl}{\relax}
\BIBdecl

\bibitem{yu2023decarbonization}
Y.~Yu, J.~Wang, Q.~Chen, J.~Urpelainen, Q.~Ding, S.~Liu, and B.~Zhang,
  ``Decarbonization efforts hindered by china’s slow progress on electricity
  market reforms,'' \emph{Nature Sustainability}, pp. 1--10, 2023.

\bibitem{lv2019intelligent}
Z.~Lv, W.~Kong, X.~Zhang, D.~Jiang, H.~Lv, and X.~Lu, ``Intelligent security
  planning for regional distributed energy internet,'' \emph{IEEE Transactions
  on Industrial Informatics}, vol.~16, no.~5, pp. 3540--3547, 2019.

\bibitem{wang2020exploring}
J.~Wang, H.~Zhong, Z.~Yang, M.~Wang, D.~M. Kammen, Z.~Liu, Z.~Ma, Q.~Xia, and
  C.~Kang, ``Exploring the trade-offs between electric heating policy and
  carbon mitigation in china,'' \emph{Nature communications}, vol.~11, no.~1,
  p. 6054, 2020.

\bibitem{WANG2023120699}
J.~Wang, Q.~An, Y.~Zhao, G.~Pan, J.~Song, Q.~Hu, and C.-W. Tan, ``Role of
  electrolytic hydrogen in smart city decarbonization in china,'' \emph{Applied
  Energy}, vol. 336, p. 120699, 2023.

\bibitem{wang2020sustainable}
J.~Wang and X.~Lu, ``Sustainable and resilient distribution systems with
  networked microgrids [point of view],'' \emph{Proceedings of the IEEE}, vol.
  108, no.~2, pp. 238--241, 2020.

\bibitem{HAKIMI2021117215}
S.~M. Hakimi, A.~Hasankhani, M.~Shafie-khah, and J.~P. Catalão, ``Stochastic
  planning of a multi-microgrid considering integration of renewable energy
  resources and real-time electricity market,'' \emph{Applied Energy}, vol.
  298, p. 117215, 2021.

\bibitem{9696007}
Z.~Zhao, J.~Guo, X.~Luo, C.~S. Lai, P.~Yang, L.~L. Lai, P.~Li, J.~M. Guerrero,
  and M.~Shahidehpour, ``Distributed robust model predictive control-based
  energy management strategy for islanded multi-microgrids considering
  uncertainty,'' \emph{IEEE Transactions on Smart Grid}, vol.~13, no.~3, pp.
  2107--2120, 2022.

\bibitem{HAN2021116830}
D.~Han and J.~H. Lee, ``Two-stage stochastic programming formulation for
  optimal design and operation of multi-microgrid system using data-based
  modeling of renewable energy sources,'' \emph{Applied Energy}, vol. 291, p.
  116830, 2021.

\bibitem{LI2021113996}
Y.~Li, C.~Wang, G.~Li, and C.~Chen, ``Optimal scheduling of integrated demand
  response-enabled integrated energy systems with uncertain renewable
  generations: A stackelberg game approach,'' \emph{Energy Conversion and
  Management}, vol. 235, p. 113996, 2021.

\bibitem{li2021coordinating}
Y.~Li, M.~Han, Z.~Yang, and G.~Li, ``Coordinating flexible demand response and
  renewable uncertainties for scheduling of community integrated energy systems
  with an electric vehicle charging station: A bi-level approach,'' \emph{IEEE
  Transactions on Sustainable Energy}, vol.~12, pp. 2321--2331, 2021.

\bibitem{peng2021hybrid}
Q.~Peng, X.~Wang, Y.~Kuang, Y.~Wang, H.~Zhao, Z.~Wang, and J.~Lyu, ``Hybrid
  energy sharing mechanism for integrated energy systems based on the
  stackelberg game,'' \emph{CSEE Journal of Power and Energy Systems}, vol.~7,
  no.~5, pp. 911--921, 2021.

\bibitem{liu2019heat}
N.~Liu, L.~Zhou, C.~Wang, X.~Yu, and X.~Ma, ``Heat-electricity coupled peak
  load shifting for multi-energy industrial parks: A stackelberg game
  approach,'' \emph{IEEE Transactions on Sustainable Energy}, vol.~11, no.~3,
  pp. 1858--1869, 2019.

\bibitem{9451164}
Z.~Qin, D.~Liu, H.~Hua, and J.~Cao, ``Privacy preserving load control of
  residential microgrid via deep reinforcement learning,'' \emph{IEEE
  Transactions on Smart Grid}, vol.~12, no.~5, pp. 4079--4089, 2021.

\bibitem{9244070}
Q.~Zhang, K.~Dehghanpour, Z.~Wang, F.~Qiu, and D.~Zhao, ``Multi-agent safe
  policy learning for power management of networked microgrids,'' \emph{IEEE
  Transactions on Smart Grid}, vol.~12, no.~2, pp. 1048--1062, 2021.

\bibitem{8331897}
E.~Foruzan, L.-K. Soh, and S.~Asgarpoor, ``Reinforcement learning approach for
  optimal distributed energy management in a microgrid,'' \emph{IEEE
  Transactions on Power Systems}, vol.~33, no.~5, pp. 5749--5758, 2018.

\bibitem{9509287}
H.~Shuai, F.~Li, H.~Pulgar-Painemal, and Y.~Xue, ``Branching dueling
  q-network-based online scheduling of a microgrid with distributed energy
  storage systems,'' \emph{IEEE Transactions on Smart Grid}, vol.~12, no.~6,
  pp. 5479--5482, 2021.

\bibitem{8839066}
X.~Lu, X.~Xiao, L.~Xiao, C.~Dai, M.~Peng, and H.~V. Poor, ``Reinforcement
  learning-based microgrid energy trading with a reduced power plant
  schedule,'' \emph{IEEE Internet of Things Journal}, vol.~6, no.~6, pp.
  10\,728--10\,737, 2019.

\bibitem{8789677}
Q.~Zhang, K.~Dehghanpour, Z.~Wang, and Q.~Huang, ``A learning-based power
  management method for networked microgrids under incomplete information,''
  \emph{IEEE Transactions on Smart Grid}, vol.~11, no.~2, pp. 1193--1204, 2020.

\bibitem{8769895}
Y.~Du and F.~Li, ``Intelligent multi-microgrid energy management based on deep
  neural network and model-free reinforcement learning,'' \emph{IEEE
  Transactions on Smart Grid}, vol.~11, no.~2, pp. 1066--1076, 2020.

\bibitem{ZHU2021117107}
Z.~Zhu, K.~{Wing Chan}, S.~Bu, B.~Zhou, and S.~Xia, ``Real-time interaction of
  active distribution network and virtual microgrids: Market paradigm and
  data-driven stakeholder behavior analysis,'' \emph{Applied Energy}, vol. 297,
  p. 117107, 2021.

\bibitem{9099902}
G.~Zhang, W.~Hu, D.~Cao, Q.~Huang, J.~Yi, Z.~Chen, and F.~Blaabjerg, ``Deep
  reinforcement learning-based approach for proportional resonance power system
  stabilizer to prevent ultra-low-frequency oscillations,'' \emph{IEEE
  Transactions on Smart Grid}, vol.~11, no.~6, pp. 5260--5272, 2020.

\bibitem{7180409}
M.~Liu and Y.~Shi, ``Model predictive control of aggregated heterogeneous
  second-order thermostatically controlled loads for ancillary services,''
  \emph{IEEE Transactions on Power Systems}, vol.~31, no.~3, pp. 1963--1971,
  2016.

\bibitem{7303968}
S.~Iacovella, F.~Ruelens, P.~Vingerhoets, B.~Claessens, and G.~Deconinck,
  ``Cluster control of heterogeneous thermostatically controlled loads using
  tracer devices,'' \emph{IEEE Transactions on Smart Grid}, vol.~8, no.~2, pp.
  528--536, 2017.

\bibitem{nakabi2021deep}
T.~A. Nakabi and P.~Toivanen, ``Deep reinforcement learning for energy
  management in a microgrid with flexible demand,'' \emph{Sustainable Energy,
  Grids and Networks}, vol.~25, p. 100413, 2021.

\bibitem{hutter2019automated}
F.~Hutter, L.~Kotthoff, and J.~Vanschoren, \emph{Automated machine learning:
  methods, systems, challenges}.\hskip 1em plus 0.5em minus 0.4em\relax
  Springer Nature, 2019.

\bibitem{awad2015machine}
M.~Awad and R.~Khanna, ``Machine learning and knowledge discovery,'' in
  \emph{Efficient learning machines}.\hskip 1em plus 0.5em minus 0.4em\relax
  Springer, 2015, pp. 19--38.

\bibitem{he2021automl}
X.~He, K.~Zhao, and X.~Chu, ``Automl: A survey of the state-of-the-art,''
  \emph{Knowledge-Based Systems}, vol. 212, p. 106622, 2021.

\bibitem{li2018metis}
Z.~L. Li, C.-J.~M. Liang, W.~He, L.~Zhu, W.~Dai, J.~Jiang, and G.~Sun, ``Metis:
  Robustly tuning tail latencies of cloud systems,'' in \emph{2018
  $\{$USENIX$\}$ Annual Technical Conference ($\{$USENIX$\}$$\{$ATC$\}$ 18)},
  2018, pp. 981--992.

\bibitem{mnih2015human}
V.~Mnih, K.~Kavukcuoglu, D.~Silver, A.~A. Rusu, J.~Veness, M.~G. Bellemare,
  A.~Graves, M.~Riedmiller, A.~K. Fidjeland, G.~Ostrovski \emph{et~al.},
  ``Human-level control through deep reinforcement learning,'' \emph{Nature},
  vol. 518, no. 7540, pp. 529--533, 2015.

\bibitem{zhang2017deeper}
S.~Zhang and R.~S. Sutton, ``A deeper look at experience replay,'' \emph{arXiv
  preprint arXiv:1712.01275}, 2017.

\bibitem{schaul2015prioritized}
T.~Schaul, J.~Quan, I.~Antonoglou, and D.~Silver, ``Prioritized experience
  replay,'' \emph{arXiv preprint arXiv:1511.05952}, 2015.

\bibitem{li2023deep2}
Y.~Li, C.~Yu, M.~Shahidehpour, T.~Yang, Z.~Zeng, and T.~Chai, ``Deep
  reinforcement learning for smart grid operations: algorithms, applications,
  and prospects,'' \emph{Proceedings of the IEEE}, vol. 111, no.~9, pp.
  1055--1096, 2023.

\end{thebibliography}
\end{document}